\documentclass[%
 reprint,
usenatbib,
 amsmath,amssymb,
 aps,
twocolumn]{aastex631}

\usepackage{graphicx}
\usepackage{dcolumn}
\usepackage{tabularx}
\usepackage{bm}

\usepackage{booktabs, tabularx, makecell, multirow, xspace}
\usepackage{graphicx}	
\usepackage{amsmath}	
\usepackage{amssymb}	
\usepackage{physics}   
\usepackage{comment}   
\usepackage{slashed}

\newcommand{\msun}{{\,\rm M_\odot}}

\newcommand{\cm}{\,{\rm cm}}

\newcommand{\K}{\,{\rm K}}

\newcommand{\pc}{\,{\rm pc}}

\renewcommand{\arraystretch}{1.8}
\newcolumntype{C}[1]{>{\centering\let\newline\\\arraybackslash\hspace{0pt}}m{#1}}

\def\aap{A\&A}

\def\apjl{ApJ}

\def\apjs{ApJS}

\newcommand{\cdm}{\texttt{CDM}}
\newcommand{\cdmnf}{\texttt{CDM-NF}}
\newcommand{\admone}{\texttt{ADM-1}}
\newcommand{\admtwo}{\texttt{ADM-2}}

\begin{document}


\title{Dissipative Dark Substructure: \\
The Consequences of Atomic Dark Matter on Milky Way Analog Subhalos}

\author[0000-0002-6505-8559]{Caleb Gemmell}
\affiliation{Department of Physics, University of Toronto, Toronto, Ontario M5S 1A7, Canada}

\author[0000-0002-7638-7454]{Sandip Roy}
\affiliation{Department of Physics, Princeton University, Princeton, NJ 08544, USA}

\author[0000-0002-6196-823X]{Xuejian Shen}
\affiliation{TAPIR, California Institute of Technology, Pasadena, CA 91125, USA}
\affiliation{Kavli Institute for Astrophysics and Space Research, Massachusetts Institute of Technology, Cambridge, MA, 02139, USA}

\author[0000-0003-0263-6195]{David Curtin}
\affiliation{Department of Physics, University of Toronto, Toronto, Ontario M5S 1A7, Canada}

\author[0000-0002-8495-8659]{Mariangela Lisanti}
\affiliation{Department of Physics, Princeton University, Princeton, NJ 08544, USA}
\affiliation{Center for Computational Astrophysics, Flatiron Institute, New York, NY 10010, USA}

\author[0000-0002-8659-3729]{Norman Murray}
\affiliation{Canadian Institute for Theoretical Astrophysics, University of Toronto, Toronto, ON M5S 3H8, Canada}

\author[0000-0003-3729-1684]{Philip F. Hopkins}
\affiliation{TAPIR, California Institute of Technology, Pasadena, CA, 91125, USA}

\begin{abstract}
Using cosmological hydrodynamical zoom-in simulations, we explore the properties of subhalos in Milky Way analogs that contain a sub-component of Atomic Dark Matter~(ADM). ADM differs from Cold Dark Matter (CDM) due to the presence of self interactions that lead to energy dissipation, analogous to Standard Model baryons. This model can arise in dark sectors that are natural and theoretically-motivated extensions to the Standard Model. 
 The simulations used in this work were carried out using \texttt{GIZMO} and utilize the FIRE-2 galaxy formation physics in the Standard Model baryonic sector. 
 For the parameter points we consider, the ADM gas cools efficiently, allowing it to collapse to the center of subhalos.  This increases a subhalo's central density and affects its orbit, with more subhalos surviving small pericentric passages. The subset of subhalos that host satellite galaxies have cuspier density profiles and smaller stellar-half-mass radii relative to CDM.
 The entire population of dwarf galaxies produced in the ADM simulations is more compact than those seen in CDM simulations,  unable to reproduce the entire diversity of observed dwarf galaxy structures.
 Additionally, we also identify a population of highly compact subhalos that consist nearly entirely of ADM and form in the central region of the host, where they can leave distinctive imprints in the baryonic disk. This work presents the first detailed exploration of subhalo properties in a strongly dissipative dark matter scenario, providing intuition for how other regions of ADM parameter space, as well as other dark sector models, would impact galactic-scale observables.
 
\end{abstract}

\section{Introduction}
\label{sec:intro}

The model of Cold Dark Matter (CDM) is theoretically minimal and has been successful at describing the large-scale structure of the Universe~\citep{Planck:2018vyg}. 
Despite these successes, there are important reasons to explore non-minimal interacting dark matter~(DM) candidates. First, discrepancies between observational data and CDM simulations have begun to appear at smaller scales---see \cite{Bullock:2017xww} for a review.
While the addition of baryons to CDM simulations can ameliorate some of the small-scale disagreements \citep{Governato:2009bg,Garrison_Kimmel_2019}, these simulations still fail to produce a population of compact dwarfs that have been found in the Local Group~\citep{Santos_Santos_2017,Jiang:2018iut,Garrison_Kimmel_2019}. This issue is also related to the \emph{Diversity Problem}, as CDM (plus baryon) simulations are unable to reproduce the wide variety of inner density slopes observed in observational data~\citep{Relatores_2019}. Second, complex dark sectors with multiple particles and forces are highly motivated from a particle theory point of view and can also ameliorate the observed small-scale tensions, e.g.,~\cite{Tulin_2018}. In particular, they are a prediction of many scenarios that solve the Hierarchy Problem, while being compatible with null results from Large Hadron Collider searches. In this work, we study an interacting dark sector on galactic scales that is motivated by both of these concerns.

In the past, most efforts to study non-minimal dark sectors have been limited to DM candidates with simple elastic self interactions, e.g.,~\cite{Tulin_2018}. However, many implementations of complex dark sectors result in inelastic scattering that  dissipate energy~\citep{Ackerman:2008kmp,Arkani_Hamed_2009, Feng:2009mn, Alves:2009nf,Cline:2013pca,Schutz:2014nka,Boddy:2016bbu,Foot:2016wvj,Agrawal:2017rvu}.
The effect of dissipative DM on small-scale structure has only recently begun to be explored in simulations \citep{Huo_2020,Shen:2021frv,Shen:2022opd}. While these initial efforts studied scenarios where dissipation is a relatively small effect, a fraction of DM can easily be highly dissipative without violating self-interaction bounds.  To this end, we explore  Atomic Dark Matter~(ADM)~\citep{Kaplan:2009de} as a case study of more strongly dissipative models.

ADM is a useful benchmark model both due to its minimality and the fact that it contains many of the most important features found in complex DM scenarios predicted by theories like the Twin Higgs~\citep{Chacko_2006, Barbieri:2016zxn, Chacko:2018vss}.
It has just a few ingredients: a dark proton $p'$ and a dark electron $e'$ with masses $m_{p'} > m_{e'}$ that are oppositely charged under a dark electromagnetic force mediated by a dark photon $\gamma'$ with coupling $\alpha'$.  The ADM makes up a fraction $f_D \equiv \Omega_\mathrm{adm}/\Omega_\mathrm{dm}$ of the DM density and gives rise to a `dark CMB' that is a factor $\xi \equiv T_\mathrm{cmb'}/T_\mathrm{cmb} < 1$ colder than the normal CMB.
In the early universe, ADM produces dark acoustic oscillations in the power spectrum, which can suppress structure formation at small scales\footnote{This small-scale suppression will also result in a larger minimum halo mass \citep{Boddy:2016bbu}, which could also be probed via the subhalo mass function.}~\citep{Cyr-Racine:2013fsa}, but current cosmological constraints still allow ADM fractions as large as $f_D \gtrsim \mathcal{O}(20\%)$~\citep{Bansal:2022qbi,Bansal:2021dfh}. Subcomponents in this range could easily have very significant astrophysical effects.

The structure of ADM on galactic scales has long been discussed, with some suggesting the possibility of a `dark disk' forming \citep{Fan_2013,Fan:2013tia,McCullough:2013jma,Randall:2014kta,Schutz:2017tfp}, or collapsed dark gas fragments distributed throughout our Galactic halo  \citep{Ghalsasi_2018,Chang_2019}. Due to the complicated dissipative and self-interacting dynamics of ADM, these discussions were based on highly simplified analytical or semi-analytical treatments. A first-principles approach was urgently needed, and this finally became possible earlier this year, when the first cosmological hydrodynamical simulations of ADM were presented \citep{Roy:2023zar}.

These simulations, which include baryons, CDM, and a 6\% fraction of ADM, focus on a Milky-Way-like host halo and demonstrate the ability of ADM to form a centrally-rotating dark disk for the two parameter points considered. 
Alongside the dark ADM gas disks, these simulations also find large populations of collapsed ADM clumps in a non-disk-like distribution, enhancing the central structure of the host halo. 
Due to the increased central densities, it is visible by eye that one of the ADM parameter points deforms the Standard Model baryonic disk. Furthermore, both ADM scenarios appear to enhance galactic rotation curves to the point of being incompatible with our Milky Way~(MW), but higher-resolution studies are required to make this definitive. 
In general, a quantitative approach is needed to understand all of the available predictions and hence extract astrophysical constraints on the ADM parameter space, from these simulations.

In this paper, we use the simulations of \cite{Roy:2023zar} to study the effects of ADM on galactic substructure, explore how ADM affects visible satellite galaxies, and understand whether such models can address some of the small-scale structure problems. 
Our investigations not only define a detailed analysis template for future simulation studies of ADM, they also demonstrate that the two ADM parameter points simulated so far appear to be in significant conflict with observational data and CDM expectations. We expect that the intuition gained from these studies will apply to broader regions of ADM parameter space, as well as other dissipative models more broadly.

The paper is organized as follows. Section~\ref{sec:simulation} reviews the details of the  simulations studied in this work. In particular, Sec.~\ref{sec:analysis} outlines the methodology used to select the subhalos and satellite galaxies in the simulated MW analogs.  Section~\ref{sec:results} analyses the properties of all DM subhalos, while Sec.~\ref{sec:sat galaxies} focuses specifically on the visible satellite galaxies. We conclude in Sec.~\ref{sec:conclusion}. Appendices~\ref{sec:resolution}--\ref{sec:error} summarize various cross checks on the analysis methods.

\section{Simulation Suite}
\label{sec:simulation}

This section briefly reviews the ADM simulations presented in \cite{Roy:2023zar}, as well as the methodology for identifying and analyzing subhalos in the simulated MW analogs.

\subsection{Zoom-In Simulations}

\begin{table*}[t]
\footnotesize
\begin{center}
\renewcommand{\arraystretch}{1.5}
\begin{tabular}{c|c|cccccccccc}
  \Xhline{3\arrayrulewidth}
\textbf{Simulation}&\textbf{Included Species} &$\mathbf{\frac{\Omega_{\rm cdm}}{\Omega_{\rm m}}}$ & $\mathbf{\frac{{m_{\rm cdm}}}{{\msun}}}$ & $\mathbf{\frac{\Omega_{\rm adm}}{\Omega_{\rm dm}}}$ &  $\mathbf{\frac{m_{\rm adm}}{\msun}}$ & $\mathbf{\frac{\Omega_{\rm b}}{\Omega_{\rm m}}}$ & $\mathbf{\frac{m_{\rm b}}{\msun}}$ & $\boldsymbol{\frac{\alpha'}{\alpha}}$ & $\mathbf{\frac{m_{p'}}{m_p}}$& $\mathbf{\frac{m_{e'}}{m_e}}$ & $\mathbf{\frac{T_{\rm cmb'}}{T_{\rm cmb}}}$\\
\hline
$\cdm$ & CDM+Bar. & 0.83 & 2.79$\times 10^{5}$ & 0 & - & 0.17 & $5.6\times10^4$ & - & - & - & -   \\
$\cdmnf$ & CDM+Bar., no FB & 0.83 & 2.79$\times 10^{5}$ & 0 & - & 0.17 &  $5.6\times10^4$ &- & - & - & -   \\
$\admone$ & CDM+ADM-1+Bar.  & 0.78 & 2.62$\times 10^{5}$ & 0.06 &  1.67$\times 10^{4}$ & 0.17 & $5.6\times10^4$ &$1/\sqrt{0.55}$ & $1.3$ & $0.55$ & 0.39  \\
$\admtwo$ &CDM+ADM-2+Bar.  & 0.78 & 2.62$\times 10^{5}$ & 0.06 &  1.67$\times 10^{4}$ & 0.17 & $5.6\times10^4$ &$2.5$ & $1.3$ & $0.55$ & 0.39  \\
  \Xhline{3\arrayrulewidth}
\end{tabular}
\end{center}
\caption{\label{tab:ADMspecies} 
Summary of the simulations from~\cite{Roy:2023zar} that are used in this work. We compare two non-ADM simulations (with and without baryonic feedback) to two with ADM. All maintain a total DM abundance of $\Omega_\mathrm{dm} = 0.83$. In ADM-inclusive simulations, ADM makes up 6\% of the DM. The symbols $m_\mathrm{cdm}, m_\mathrm{adm}$, and $m_\mathrm{b}$ represent particle masses for various simulation components. ADM's cooling physics is defined by the dark fine-structure constant~($\alpha'$), dark proton mass~($m_{p'}$), dark electron mass~($m_{e'}$), and dark CMB temperature $T_\mathrm{cmb'}$ which we list relative to the Standard Model values (unprimed quantities). A higher $\alpha'$ in $\admtwo$ leads to faster ADM cooling and a greater dark cut-off temperature due to the increased dark binding energy. Thus, the dark gas cools more efficiently in $\admtwo$ than in $\admone$, but predominantly at greater temperatures ($T \gtrsim 10^5 K$).
}
\end{table*}

The dataset includes four cosmological hydrodynamical zoom-in simulations of MW analogs. The matter component of the $\cdm$ and $\cdmnf$ simulations is comprised of CDM and baryons. Between the two, $\cdm$ includes baryonic stellar feedback while $\cdmnf$ does not. The remaining two simulations, denoted by $\admone$ and $\admtwo$, contain CDM, baryons (with feedback), and ADM, a subcomponent of the DM. Along with the specified ADM mass fraction  ($\Omega_{\rm{adm}}/\Omega_{\rm{dm}}$), the \texttt{ADM} simulations are dependent on three additional parameters: the dark fine-structure constant~($\alpha'$), dark proton mass~($m_{p'}$), and dark electron mass~($m_{e'}$). The main features of the simulation suite are summarized in this section and Tab.~\ref{tab:ADMspecies}; further details can be found in \cite{Roy:2023zar}.

All simulations utilize the code~\texttt{GIZMO}~\citep{Hopkins2015}. The baryonic physics adopts the prescription from the Feedback In Realistic Environments~(FIRE) project~\citep{Hopkins2014}, specifically the ``FIRE-2'' version~\citep{Hopkins2017b,Hopkins2018}, including low-temperature gas cooling~\citep{Ferland1998-CLOUDY, Wiersma2009}, heating from a meta-galactic radiation background~\citep{Onorbe2016,FG2020} and stellar sources, star formation, as well as explicit models for stellar and supernovae feedback \citep{Hopkins2014}. The star formation criteria for baryonic gas include reaching a critical density of $\sim1000~\cm^{-3}$, hosting non-zero molecular fractions \citep{Krumholz2011}, and becoming Jeans-unstable and locally self-gravitating \citep{Hopkins2013}. For $\cdmnf$ only, we remove all forms of stellar, supernovae, and radiative feedback.

ADM is implemented in \texttt{GIZMO} as a separate gas species that is decoupled hydrodynamically from the baryonic gas. The ADM cooling functions are implemented as described in \cite{Roy:2023zar} and \cite{Rosenberg2017}, assuming no meta-galactic dark radiation background in the ADM sector except for the dark Cosmic Microwave Background~(CMB). The cooling functions assume ionisation-recombination equilibrium for the ADM gas and do not include dark molecular cooling~\citep{Ryan:2021dis}. Dark molecular cooling is relevant for simulating dense regions of ADM gas at scales below the current resolution of the simulations.

There are several key features of ADM atomic cooling.  At temperatures far above the binding energy of the dark hydrogen ($B_0' = \frac{1}{2} \alpha'^2 m_{e'} c^2 $), Bremsstrahlung cooling is dominant and monotonically increases with temperature.  At temperatures closer to $B_0'$, there is a prominent collisional excitation cooling peak associated with non-zero neutral dark hydrogen fractions.  Lastly, there is a sharp cut-off in the atomic cooling rates at temperatures significantly below $B_0'$ where the dark gas is neutral; we refer to this transition as the cut-off temperature. Greater $B_0'$ implies greater pressure support for dense, cold gas in the center of halos. Because both $B_0'$ and the cooling rate increase with $\alpha'$, the ADM gas in $\admtwo$ cools more efficiently than that of $\admone$, but at greater temperatures ($T \gtrsim 10^5 K$).

Once ADM gas cells become locally self-gravitating and Jeans-unstable, they are turned into ``clump'' particles over the free-fall time scale. These clump particles have the same mass as the ADM gas cell ($1.67\times10^4\msun$) and represent collisionless objects at the resolution scale of the simulation. This is due to the ADM gas cloud collapsing and fragmenting into dark compact objects~\citep{Gurian:2022nbx} at scales smaller than we are simulating, such as black holes or dark stars. Once formed, the clump particles interact only gravitationally like the CDM particles.

The force softening for both the baryonic and ADM gas particles uses the algorithm from~\citet{Price2007}.
The minimum gas softening is $h_{\rm gas} = 1.4\pc$ while the CDM force resolution of the simulations is $h_{\rm dm} = 40\pc$. 

When considering the inner regions of substructure, we must verify that the properties of the halo have converged and are resolution-independent. A commonly used metric is the convergence radius criterion specified by \cite{Power:2002sw}, mandating that the $N$-body relaxation time in DM-only simulations, $t_{\rm relax}(r)$, is shorter than the age of the universe, $t_0$. This is determined by the equation,
\begin{equation}
    \frac{t_{\rm{relax}}(r)}{t_0} = \frac{\sqrt{200}}{8}\frac{N}{\rm{ln}\:N}\bigg(\frac{\overline{\rho}(r)}{\rho_{\rm{crit}}}\bigg)^{-1/2} \, ,
\end{equation}
where $N$ is the number of particles enclosed by $r$, $\overline{\rho}(r)$ is the mean density enclosed by $r$, and $\rho_{\rm{crit}}$ is the critical density of the universe. \cite{Hopkins:2017ycn} conducted an extensive study to apply the convergence radius criterion to simulations with FIRE-2 baryonic physics and found the DM to have converged for radii larger than $r^{\rm{conv}}_{\rm{ \scriptscriptstyle{DM}}}$, defined by the relation $t_{\rm{relax}}(r^{\rm{conv}}_{\rm \scriptscriptstyle{DM}})=0.06\:t_0$. This is approximately the radius that encloses 200 CDM particles. However, in the same study, this was also shown to be a conservative estimate and that simulations with baryons could be resolved to much smaller radii. The effect of ADM on DM convergence is currently unknown, but the convergence of the ADM structure theoretically should be much smaller as it is determined by the minimum gas softening~(1.4~pc) and not the CDM force resolution~(40~pc). To incorporate the effect of ADM on the convergence radii for all collisionless DM in the \texttt{ADM} simulations, we calculate $r^{\rm{conv}}_{\rm{ \scriptscriptstyle{DM}}}$ using both CDM and ADM clump particles. Thus, in the regime that ADM is able to cool and form dense regions in the center of subhalos, $r^{\rm{conv}}_{\rm{ \scriptscriptstyle{DM}}}$ is much smaller than the CDM-only expectation, due to the high number of ADM clumps at small radii. As a further check for convergence, we compare our results with low-resolution simulations in Appendix~\ref{sec:resolution} and comment on what trends are resolution-independent.

The primary halo in the zoom-in area that forms the focus of this work is \texttt{m12i}. It was selected from the standard FIRE-2 suite~\citep{Wetzel2022,Hopkins2014} and has a CDM mass of $\sim 10^{12}\msun$, a stellar disk mass of $7.4\times10^{10}\msun$, as well as a stable merger history for $\cdm$ below $z < 1$~\citep{Hopkins2018}. The publicly released high-resolution \texttt{m12i} simulation \citep{Wetzel:2022man} is referred to throughout this study as a cross-check against the $\cdm$ simulation, as it has approximately a factor of eight better mass resolution ($m_{\rm{cdm}} = 3.5\times10^4 \msun$) than the $\cdm$ simulation.

The initial transfer functions for the baryons, CDM, and ADM particles are calculated using a modified version of \texttt{CLASS}~\citep{Blas2011,Bansal:2021dfh,Bansal:2022qbi}, which accounts for the cosmological history of the ADM. \texttt{MUSIC}~\citep{Hahn2011} is then used to generate the initial conditions at $z\sim 100$. 

\newpage

\subsection{Subhalo Identification}
\label{sec:analysis}

To identify DM subhalos in our MW analogs, we use the \texttt{ROCKSTAR} 6D halo finder.\footnote{Specifically, we use the adapted version of \cite{2020ascl.soft02014W}, found at \url{https://bitbucket.org/awetzel/rockstar-galaxies/}.} In CDM-only simulations, \texttt{ROCKSTAR} is run only on the DM particles for numerical stability~\citep{Samuel:2019ylk}. Equivalently, we run our analysis on the CDM particles and collisionless ADM clumps within 300~kpc of the host's center. The friends-of-friends linking percentage in \texttt{Rockstar} is typically set to 70\%, but for this study, we conservatively increase it to 80\%. This choice is inspired by the fact that ADM can cool and fragment, potentially leading to substructure forming at smaller scales in the \texttt{ADM} simulations. Increasing the linking percentage ensures that the clustering algorithm is not dominated by the dense regions of ADM clumps. However, the effect of changing this parameter is very minor, see Appendix~\ref{sec:rockstar}. Once identified, subhalos are classified by $R_{200, {\rm m}}$, the radius that encloses 200 times the mean matter density, and $M_{200, {\rm m}}$, the total DM mass contained within this radius.

In CDM simulations, a cut is typically applied on particle numbers to ensure that spurious subhalos comprising only a few simulation particles are not considered. 
However, for the \texttt{ADM} simulations, the subhalos consist of both CDM particles ($m_{\rm cdm} = 2.62\times10^5\msun$) and ADM clumps of mass $1.67\times10^4\msun$---an order-of-magnitude smaller than the CDM particles. In these simulations, there may be subhalos where a small number of CDM particles dominate the total mass of the subhalo. For example, a subhalo of 2~CDM particles and 49~ADM clumps may pass a particle number cut, $N>50$, but 39\% of the subhalo's mass/gravitational potential is determined by the two CDM particles. To account for such scenarios, we assume that the number cut in CDM-only simulations corresponds to requiring the relative uncertainty on the halo's mass to be below some threshold. Assuming that this relative uncertainty scales as $N^{-p/2}$ in the small-number-limit, where $p$ is some \emph{a priori} unknown scaling constant, generalizing this threshold to include ADM particles corresponds to adding the different particle species' contribution in quadrature:
\begin{equation}
    \sigma = \Biggr[\bigg(\frac{f_{\rm{adm}}}{N_{\rm{adm}}^{p/2}}\bigg)^2 + \bigg(\frac{f_{\rm{cdm}}}{N_{\rm{cdm}}^{p/2}}\bigg)^2\Biggr]^{1/2p} 
     < \frac{1}{\sqrt{N_{\rm{cut}}}}\, ,
     \label{eq:threshold}
\end{equation}
where $f_{\rm{adm}}$ ($f_{\rm{cdm}}$) is the subhalo ADM~(CDM) mass fraction, $N_{\rm{adm}}$~($N_{\rm{cdm}}$) is  the number of ADM~(CDM) particles contained within the subhalo, and $N_{\rm cut}$ is the number count that the halo must satisfy. An overall exponential factor of $1/p$ is included to remove any explicit $p$ dependence from the final threshold value.

The threshold $\sigma$ should recover the standard particle number cut, $N > N_{\rm{cut}}$, in the limit where the halo contains only one particle type, as well as the case where both particle types have the same mass. This uniquely sets $p = 1$ in Eq.~(\ref{eq:threshold}), which also results in the cut being more conservative than just $N>N_{\rm{cut}}$ for subhalos containing simulation particles of different masses, which is sensible.

In this work, we set $N_{\rm{cut}} = 50$. This corresponds to the halo mass in the $\cdm$ simulation where the subhalo mass function begins to diverge from that of the high-resolution \texttt{m12i} simulation, shown in Fig.~\ref{fig:SHMF_no_cut}. This establishes two different mass thresholds. For the $\cdm$ and $\cdmnf$ simulations, all subhalos have masses larger than $50\times m_{\rm{cdm}}=1.4\times10^7\msun$. For $\admone$ and $\admtwo$, all subhalos have masses larger than $50\times m_{\rm{adm}}=8.4\times10^5\msun$. In Appendix~\ref{sec:error}, we compare our results to those using a larger threshold, $N>200$. 

To identify visible satellite galaxies, we use \texttt{HaloAnalysis} \citep{2020ascl.soft02014W} to associate stellar particles to DM subhalos identified with \texttt{Rockstar} following the post-processing steps of~\cite{Samuel:2019ylk}. For each subhalo, an initial position cut is applied so that we only consider stellar particles within $0.8 \times R_{\rm 200, m}$ of the subhalo's center, out to a maximum of 30~kpc.
Additionally, the velocity of each star particle relative to the subhalo must satisfy $v \leq 2 \times V_{\rm{circ}}^{\rm{max}}$ (maximum circular velocity) and $v \leq 2 \times \sigma_{\rm v,dm}^{\rm 3D}$ (3D velocity dispersion of DM). Following this step, an iterative cut is applied, keeping only stellar particles with $r < 1.5 \times r_{90}$ (radius enclosing 90\% of stellar particles in the subhalo) and $v < 2 \times \sigma_{\rm v,*}^{\rm 3D}$ (3D velocity dispersion of subhalo stellar particles), until the number of stellar particles converges at the 1\% level. Once baryon-containing subhalos are identified, we only include subhalos with at least 10 stellar particles and an average stellar density greater than $300 \msun $ kpc$^{-3}$, which results in an overall stellar mass cut, M$_{*}>10^{5.5}\msun$.

\section{General Subhalo Properties}
\label{sec:results}

In the MW analogs for $\cdm$, $\cdmnf$, $\admone$, and $\admtwo$, a total of 455, 388, 523, and 264 subhalos, respectively, are identified. Of the subhalos in the \texttt{ADM} simulations, 42\%~(13\%) contain ADM clumps, in addition to CDM particles, for $\admone$~($\admtwo$). In Sec.~\ref{sec:ADM subhalos}--\ref{sec:orbits}, we present the subhalo mass and radial distributions for each simulation; investigate the internal mass distribution of the subhalos; study the properties of their orbits around the host halo; and discuss how that may influence the total number of subhalos as well as their radial distribution. 
Key quantities include the subhalo density profiles, compactness, and orbital parameters. The compactness is defined by $R_{200, {\rm m}}/R_{1/2}$, where $R_{1/2}$ is the radius that encloses 50\% of the subhalo's total DM mass. 
Finally, we briefly discuss ADM overdensities in the MW  disk in Sec.~\ref{sec:clumps}.

\subsection{Density Profiles}
\label{sec:ADM subhalos}

\begin{figure}
    \centering
    \includegraphics[width=\linewidth]{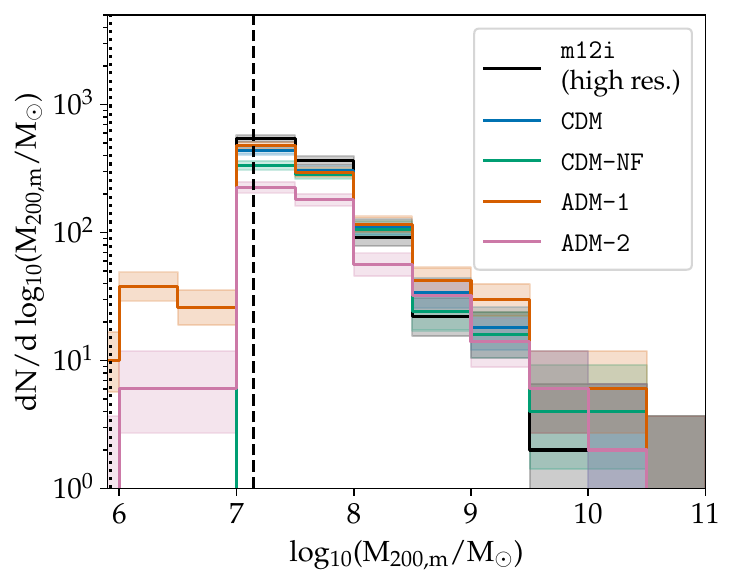}
    \caption{The subhalo mass function at $z=0$ for the $\cdm$, $\cdmnf$, $\admone$, and $\admtwo$ simulations.  The shaded bands indicate the 1$\sigma$ Poissonian error for the count in each bin. The vertical dashed-black line indicates the mass-quality threshold for the $\cdm$ and $\cdmnf$ simulations, $50 \times m_{\rm cdm} = 1.4\times10^{7} \msun$, while the vertical dotted-black line indicates the mass-quality threshold for the $\admone$ and $\admtwo$ simulations, $50 \times m_{\rm adm} = 8.4\times10^{5} \msun$. The solid black is the subhalo mass function obtained from the  public FIRE-2 data for the \texttt{m12i} simulation \citep{Wetzel_2023}. In the large mass bins (M$_{200,\rm{m}} > 10^{9.5}\msun$), the subhalo mass function for $\cdm$ is the exact same as $\admtwo$ and, similarly, so is the subhalo mass function for \texttt{m12i} for M$_{200,\rm{m}}> 10^{10}\msun$. All simulations except $\admtwo$ agree within the uncertainty for the highest mass bins, while $\admtwo$ shows an obvious depletion across multiple mass bins. There is an additional population of light \texttt{ADM} subhalos resolved below the $\cdm
    $ mass threshold that consists nearly entirely of ADM clumps ($\sim 98\%$ ADM mass fraction).}
    \label{fig:SHMF}
\end{figure}

Figure \ref{fig:SHMF} shows the subhalo mass function at $z=0$ for each simulation in the suite: $\cdm$~(blue), $\cdmnf$~(green), $\admone$~(orange), and $\admtwo$~(pink). The shaded bands correspond to the 1$\sigma$ Poissonian error in counts for each bin. For comparison, the solid black line shows the subhalo mass function obtained from the public FIRE-2 data for the $\texttt{m12i}$ simulation~\citep{Wetzel_2023}, for subhalos with $M_{200,\rm{m}} > 1.4\times10^{7} \msun$, which is the $\cdm$ mass threshold indicated by the black dashed line. The \texttt{ADM} mass threshold is indicated by the black dotted line on the left edge of the plot.

\begin{figure*}
    \centering
    \includegraphics[width=\linewidth]{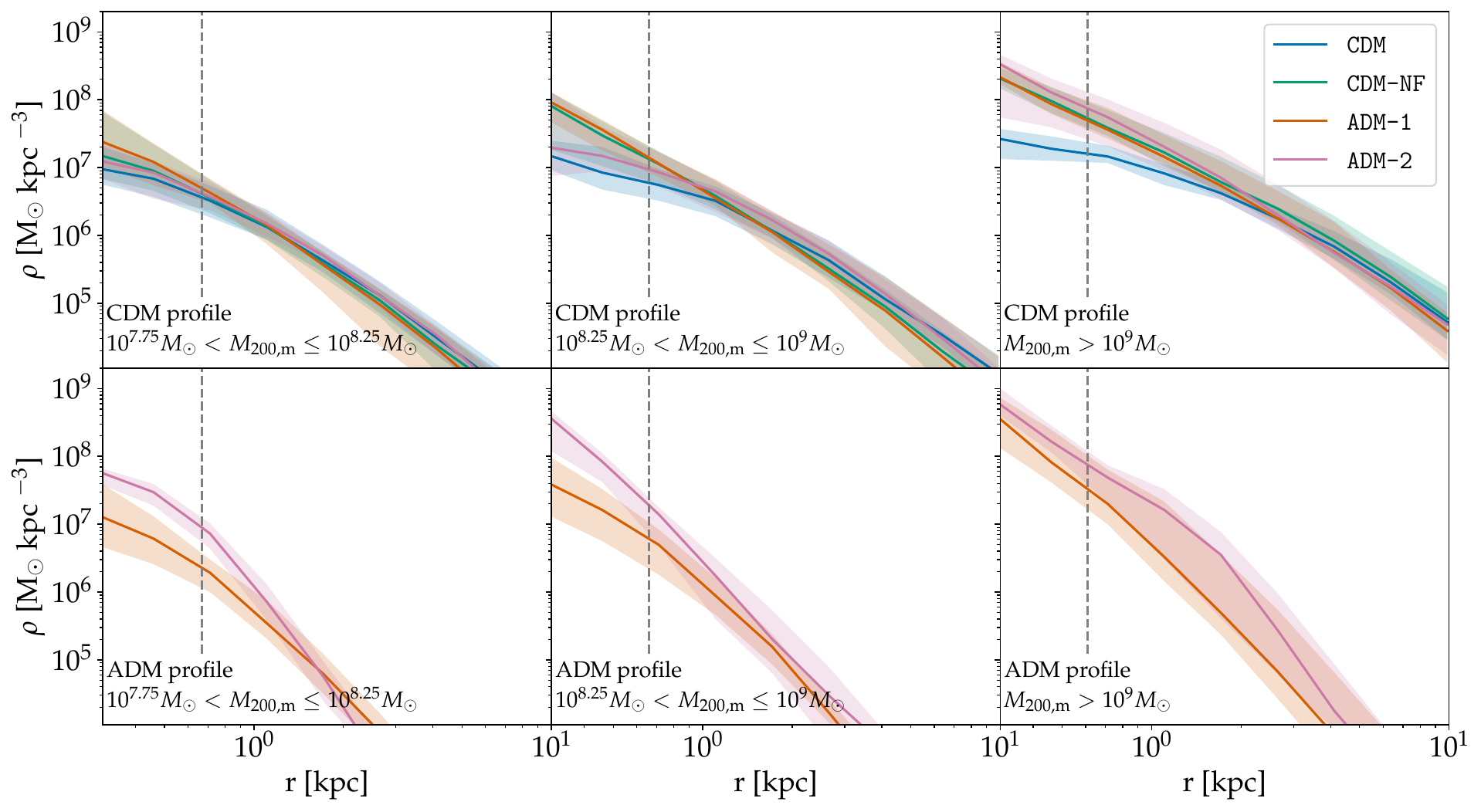}
    \caption{ \textbf{Top:} Median CDM density profiles across subhalos, per radial bin. Colored bands indicate the 68\% containment regions. Median and percentile values are calculated for subhalos with halo masses in various ranges: $10^{7.75} \msun < M_{200, {\rm m}} \leq 10^{8.25} \msun$~(left), $10^{8.25} \msun < M_{200, {\rm m}} \leq 10^9 \msun$~(middle), and $M_{200,{\rm m}} > 10^9 \msun$~(right). The dashed gray line depicts the largest convergence radius, $r_{\rm{DM}}^{\rm{conv}}$, for each mass range. For $r > r_{\rm{DM}}^{\rm{conv}}$, the CDM density profiles align for the lowest-mass subhalos, but the \texttt{ADM} and $\cdmnf$ profiles show an enhanced central density for the more massive subhalos. Because $\cdmnf$ has no baryonic feedback, the subhalos are not cored relative to $\cdm$, with adiabatic contraction also contributing to the enhanced central density for the largest mass range. The subhalos in the \texttt{ADM} simulation likely undergo dark adiabatic contraction due to the cooling ADM gas. \textbf{Bottom:} Median ADM (clumps + gas) density profiles across ADM-containing subhalos, per radial bin. The total DM profiles for \texttt{ADM} are enhanced because the ADM gas cools and forms ADM clumps at the center of a subhalo, which then contributes to the total DM density profile.}
    \label{fig:density_triptych}
\end{figure*}

Comparing simulations above the $\cdm$ mass threshold, the largest deviations are in the lower mass bins, close to the threshold where resolution effects become important. For heavier masses, there is good agreement within the count distribution for all simulations except $\admtwo$, whose mass function is suppressed across multiple bins. This behavior will be explained later in this section. Below the $\cdm$ mass threshold, only the \texttt{ADM} simulations have some resolved subhalos, due to the order-of-magnitude lighter ADM clump particles. We emphasize this is not the complete subhalo mass function below the $\cdm$ mass threshold, as a higher-resolution \texttt{ADM} simulation would also contain subhalos with a majority of CDM particles down to these masses. Yet, the individual subhalos consisting of the light ADM particles are themselves significant, as we explore further below.

In Fig.~\ref{fig:density_triptych}, the subhalo radial DM density profiles are plotted for each simulation; these profiles correspond to CDM~(top row), as well as the ADM clumps and gas~(bottom row) for the \texttt{ADM} simulations. Each panel in the figure corresponds to a different range in subhalo $M_{200, {\rm m}}$: from left to right, $10^{7.75}$--$10^{8.25}~\msun$, $10^{8.25}$--$10^9~\msun$, and $> 10^9~\msun$. To obtain the profiles, each subhalo's density is calculated in 10 logarithmically-spaced bins from 0.25~kpc to 12~kpc. The median density, represented by the solid lines, is calculated for each radial bin across all subhalos in that mass range and simulation. The shaded bands represent the 68\% containment in densities. Conservatively, the largest subhalo convergence radius for each plot is indicated by the gray dashed line, which is always from the $\cdm$ simulation.

Comparing the CDM density profiles for radii larger than $r_{\rm{DM}}^{\rm{conv}}$, the profiles are consistent between simulations for the smallest subhalo masses. For the intermediate subhalo masses, the profiles start to deviate at radii just above $r_{\rm{DM}}^{\rm{conv}}$. There is a possible enhancement in the \texttt{ADM} and $\cdmnf$ simulations compared to $\cdm$, but the large spread in the subhalo density profiles from each simulation makes it difficult to determine the significance of this. However, for the largest mass subhalos, the deviations from $\cdm$ become more pronounced.

There are two conflating factors that may be causing the inner-density enhancement in $\cdmnf$ subhalos relative to $\cdm$.  Firstly, $\cdmnf$ has no baryonic feedback and thus its subhalos are not cored relative to those in $\cdm$, especially for the largest masses. Secondly, the $\cdmnf$ subhaloes typically have larger stellar-to-halo mass ratios ($M_*/M_{200,\rm{m}}\sim0.1$) than those in $\cdm$. While in $\cdm$ adiabatic contraction~\citep{Gnedin:2004cx,Chan:2015tna} is not an influential effect for subhalos with $M_{\rm 200, m} \lesssim 10^9 \msun$, the $\cdmnf$ profiles in this range can actually be enhanced due to the lack of feedback.

While we have discussed the $\cdm$ profile above the conservative limit $r_{\rm{DM}}^{\rm{conv}}$, the \texttt{ADM} subhalos can be resolved down to smaller radii due to the large number of ADM clumps that form and occupy their central regions. For example, central enhancements are likely converged even within $r\lesssim0.3$ kpc for the most massive subhalos, as they contain an average of $1.6\times10^4$ ($2.7\times10^4$) ADM clumps for $\admone$ ($\admtwo$).

For the intermediate-mass subhalos, the $\admone$ density profiles are enhanced much like in $\cdmnf$. Even though the standard baryonic feedback is present in the $\admone$ simulation, the cooling ADM gas is likely providing an additional `dark baryonic contraction' effect that contracts the CDM profile, making it cuspier than the \texttt{CDM} expectation. $\admtwo$ does not show the same enhancement, likely because a smaller fraction~(28\%) of the subhalos contain ADM clumps compared to $\admone$~(89\%) in this mass range. This is due to the fact that the cooling rate for $\admtwo$ only becomes efficient at higher temperatures/virial masses due to its larger cut-off temperature compared to $\admone$, so the dark adiabatic contraction has a smaller effect for the intermediate-mass subhalos.
This contraction in the CDM profiles was also noted in \cite{Roy:2023zar}. The central CDM densities of the host galaxies ($<1$ kpc) for both \texttt{ADM} simulations have a $\gtrsim40\%$ enhancement compared to $\cdm$.

Indeed, $\admtwo$ shows a similar central-density enhancement as $\admone$ for the most massive subhalos, where a majority (73\%) of its subhalos contain ADM clumps, signifying the ADM gas is now able to cool and collapse efficiently. For this mass range, 100\% of $\admone$ subhalos contain ADM clumps with an average total ADM clump mass of $\sim10^{8.7}\msun$, while 73\% of $\admtwo$ subhalos contain ADM clumps with an average total ADM clump mass of $\sim10^{8.9}\msun$. These large central ADM clump densities also contribute to the density profile enhancement in the centers of subhalos.

To summarize, the inner subhalo density profiles in $\cdmnf$ can be enhanced compared to those in $\cdm$, which is to be expected as there is no baryonic feedback to core the subhalos. Additionally, due to the subhalos forming with large stellar-to-halo mass fractions, baryonic contraction could also enhance the central densities. The ADM-containing subhalos in the \texttt{ADM} simulations also have enhanced central DM densities compared to the $\cdm$ simulation. The ADM gas then collapses into central ADM clumps, contributing to the total DM density profile. Additionally, the cooling ADM gas back-reacts on the CDM particles, resulting in a dark adiabatic contraction that leads to cuspier inner density profiles. Between $\admone$ and $\admtwo$, $\admone$ is able to cool at smaller virial masses, so this effect is present for a larger population of lighter subhalos compared to $\admtwo$. However, at larger subhalo masses, $\admtwo$ cools more efficiently, forming larger central ADM masses in the subhalos containing ADM.

\begin{figure*}
    \centering
    \includegraphics[width=\linewidth]{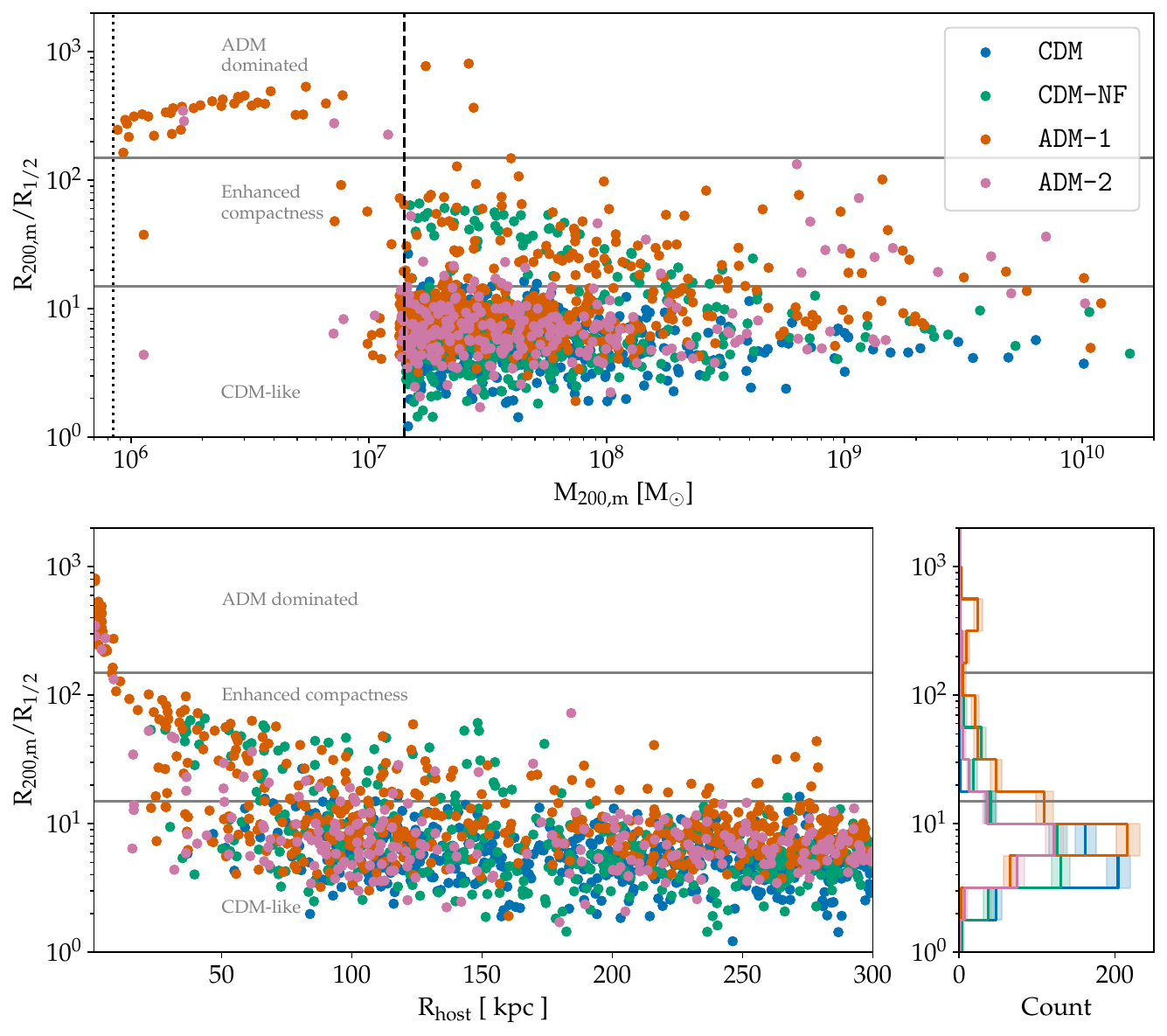}
    \caption{Distributions of subhalo compactness, as defined by $R_{200, {\rm m}}/R_{1/2}$, for $\cdm$~(blue), $\cdmnf$~(green), $\admone$~(orange), and $\admtwo$~(pink) simulations. \textbf{Top:} Plot of $R_{200, {\rm m}}/R_{1/2}$ versus halo mass, $M_{200, {\rm m}}$. The vertical dashed black line represents the $\cdm$ mass threshold, $50 \times m_{\rm cdm} = 1.4\times10^{7} \msun$, while the vertical dotted black line represents the \texttt{ADM} mass threshold, $50 \times m_{\rm adm} = 8.4\times10^{5} \msun$. The majority of subhalos have $R_{200, {\rm m}}/R_{1/2} < 15$, which we classify as `CDM-like', as this includes 99\% of the $\cdm$ subhalos. There is also a large population of \texttt{ADM} subhalos with $R_{200, {\rm m}}/R_{1/2} > 15$, with compactness relatively enhanced compared to the majority of $\cdm$ subhalos.  These correspond to \texttt{ADM} subhalos that contain a larger average mass fraction of ADM clumps in $\admone$ $(28\%)$ and $\admtwo$ $(36\%)$. The subhalos that are mostly comprised of ADM clumps, which we refer to as `ADM-dominated,' form even more compact subhalos with $R_{200, {\rm m}}/R_{1/2} > 150$. \textbf{Bottom left:} Plot of $R_{200,{\rm m}}/R_{1/2}$ versus distance to host galaxy, $R_{\rm host}$. The subhalos with enhanced compactness appear more likely to occupy the inner regions of the galaxy. Additionally, the most compact subhalos are all found within 10~kpc of the galactic center. \textbf{Bottom right:} Histogram of $R_{200, {\rm m}}/R_{1/2}$ values.  All simulations peak in the `CDM-like' region, but the \texttt{ADM} simulations are peaked at slightly larger values of compactness and have more extensive tails. The Enhanced-compactness region is mainly occupied by a second smaller peak for $\cdmnf$ and the extended tails of the \texttt{ADM} simulations. The ADM-dominated region is mostly occupied by a second peak of $\admone$ subhalos.} 
    \label{fig:rhalf}
\end{figure*}

\subsection{Compactness}
\label{sec:compactness}

Next, we study the influence of ADM on a subhalo's compactness, as quantified by R$_{200,\rm{m}}/$R$_{1/2}$. Larger values of compactness signify that a greater amount of a subhalo's mass is concentrated in its central region. The top panel of Fig.~\ref{fig:rhalf} shows how a subhalo's compactness is related to its mass. In this plot, there are three main regions populated by the subhalos. The first region corresponds to $R_{200, {\rm m}}/R_{1/2}\lesssim15$.  This is where the majority~(99\%) of subhalos in the $\cdm$ simulation~(blue) are clustered. A significant fraction of the subhalos in $\cdmnf$~(green), $\admone$~(orange), and $\admtwo$~(pink) occupy this same region: 85\%, 72\%, and 90\%, respectively. Importantly, the majority of these subhalos do not contain any ADM clumps in the $\admone$~(77\%) and $\admtwo$~(95\%) simulations. For this reason, we classify these subhalos as `CDM-like.'

The second region corresponds to $R_{200, {\rm m}}/R_{1/2}$ in the $\sim 15$--150 range and is populated by a subset of subhalos in the $\cdmnf$~(15\%), $\admone$~(21\%), and $\admtwo$~(9\%) simulations.  We refer to these subhalos as `Enhanced compactness.'
The small fraction of $\cdmnf$ subhalos in this region is due to the fact that the lack of feedback leads to more concentrated baryonic distributions, which in turn further concentrates the DM halo. Of the subhalos from $\admone$ and $\admtwo$ in this region, a larger fraction contain ADM clumps (88\% and 74\%, respectively) compared to those in the CDM-like region; the average ADM mass fraction of these subhalos is 0.28 and 0.36, respectively, compared to the vanishingly small fractions for the CDM-like subhalos. In general, the more ADM is present in a subhalo, the more compact the subhalo's total DM distribution will be.  Thus, the enhanced compactness for the \texttt{ADM} subhalos is due to the additional contribution of ADM clumps in their central regions, and also dark adiabatic contraction. Moreover, the fact that the $\admtwo$ gas cools more efficiently in more massive subhalos explains why the pink $\admtwo$ points in this region occur predominantly at larger masses (M$_{200,\rm{m}}>5\times10^{8}\msun$).  

The last region in the top panel of Fig.~\ref{fig:rhalf} corresponds to a small fraction of subhalos unique to  $\admone$ (7\%) and $\admtwo$ (2\%). These subhalos are extremely compact, with $R_{200, {\rm m}}/R_{1/2}\gtrsim 150$. The subhalos in both simulations have a large average ADM clump mass fraction of $\sim 98\%$, allowing them to be resolved at masses below the $\cdm$ threshold. We refer to these subhalos as `ADM dominated' and their origin is discussed in Sec.~\ref{sec:clumps}.

The bottom panel of Fig.~\ref{fig:rhalf} shows how a subhalo's compactness is correlated with its distance to the center of the host galaxy, $R_{\rm host}$. In general, the more concentrated the subhalo, the closer it can be found to the center of its host.  For example, 22\%, 43\%, and 35\% of subhalos below $R_{\rm host} \sim 50$~kpc belong to the CDM-like, Enhanced-compactness, and ADM-dominated populations, respectively. This is likely due to the fact that subhalos with increased central densities are more likely to survive passages close to the host center and thus occupy closer orbits~\citep{Errani_2016}, which could explain why there are almost twice as many Enhanced-compactness subhaloes within 50 kpc of the host as CDM-like subhalos. However, tidal stripping of outer material on these close orbits could also contribute to the increased compactness of these subhalos, thus it is unclear how much of the enhancement in compactness is due to the increased central density due to ADM / lack of feedback or depletion of outer material due to tidal stripping. We will disambiguate the relation between compactness and radial distance in Sec.~\ref{sec:orbits} by studying the subhalo orbital parameters.

The population of extremely compact, ADM-dominated subhalos is mostly within $\sim 10$~kpc of the galactic center, in the neighbourhood of the disk. This suggests that these subhalos are not infalling into the MW-like host, but instead form in the disk due to the fragmenting ADM gas, which we discuss in Sec.~\ref{sec:clumps}.

\subsection{Radial Distribution and Orbits}
\label{sec:orbits}

\begin{figure}
    \centering
    \includegraphics[width=\linewidth]{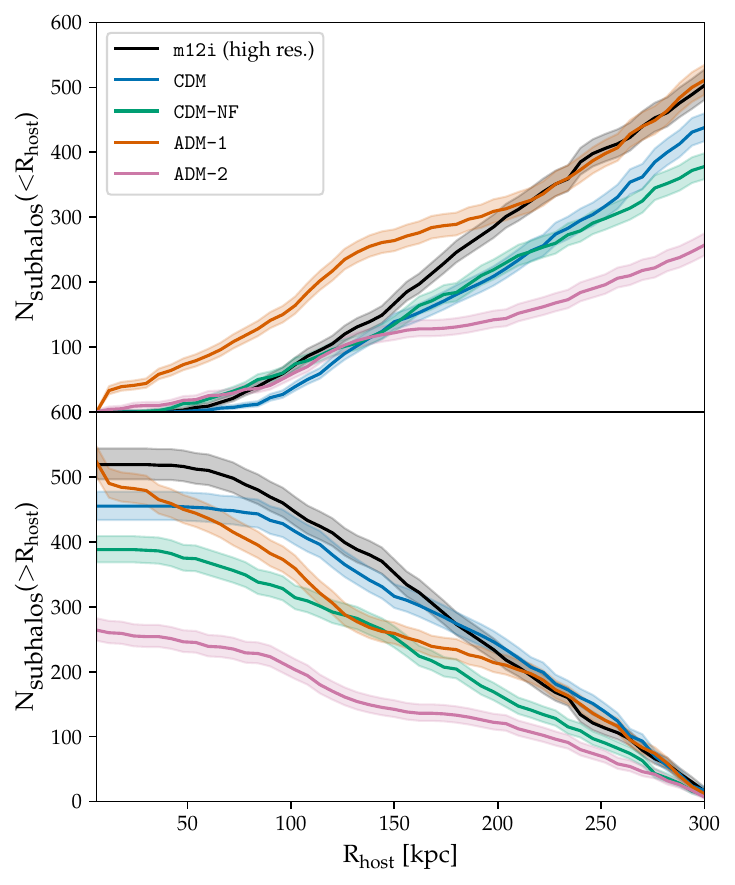}
\caption{Cumulative distributions of subhalos at $z=0$.  Results are shown for the $\cdm$~(blue), $\cdmnf$~(green), $\admone$~(orange), and $\admtwo$~(pink) simulations. Shaded bands indicate the 1$\sigma$ Poissonian error in the count. The solid black is the cumulative distribution obtained from the  high-resolution~(high res.) public FIRE-2 data for the \texttt{m12i} simulation \citep{Wetzel_2023}. \textbf{Top:} Cumulative number of subhalos within distance to host galaxy, R$_{\rm host}$. Both \texttt{ADM} simulations contain a larger fraction of subhalos within the inner $\sim 125$~kpc of the galaxy, compared to the CDM-only simulations. The total number of subhalos is enhanced in $\admone$, but depleted in $\admtwo$.
\textbf{Bottom:} Cumulative number of subhalos outside of distance to host galaxy, R$_{\rm host}$. There is good agreement between \texttt{m12i} and $\cdm$ within uncertainty outside of $\sim 100$ kpc, with $\admone$ also agreeing outside of $\sim 200$ kpc.}
\label{fig:radial_dist}
\end{figure}

Figure~\ref{fig:radial_dist} shows the cumulative radial distribution for subhalos within a given radius and summarizes many of the trends previously discussed.
The $\cdm$ distribution~(blue) has a similar shape to the high-resolution \texttt{m12i} distribution~(black), but fewer subhalos, which is suggestive of a resolution effect. Indeed, if we increase the particle number cut to a larger mass ($\sim 200 \times m_{\rm{cdm}}$, shown in Fig.~\ref{fig:radial_200}), the two simulations agree within the expected spread---thus, the difference in Fig.~\ref{fig:radial_dist} is mostly due to low-mass subhalos that are not resolved in $\cdm$. Relative to $\cdm$, the $\cdmnf$ distribution~(green) has fewer total subhalos, but they are positioned closer to the galactic center.

The radial distribution of subhalos in both $\admone$ and $\admtwo$ are quite distinctive from the $\cdm$ case. Firstly, we note that $\admone$ shows a slight total enhancement compared to $\cdm$, while the $\admtwo$ simulation has a large relative depletion in the total number of subhalos. This depletion was first pointed out in Fig.~\ref{fig:SHMF} for $M_{200,\rm{m}}<10^{8.5}\msun$, and we will explore it later in this section. Secondly, the cumulative distributions of both \texttt{ADM} simulations flatten in the range 125--200~kpc, a feature not present in the $\cdm$ or $\cdmnf$ distributions.
This feature can also be thought of as an enhancement in the fraction of subhalos at radii $\lesssim125$ kpc. Comparing the ratio $N(r<125~\text{kpc})/N(r <300~\text{kpc})$, we find $\admone$~(0.47) and $\admtwo$~(0.42) have comparable values that are larger than those for $\cdm$~(0.22) and $\cdmnf$~(0.27).

\begin{figure*}
    \centering
    \includegraphics[width=\linewidth]{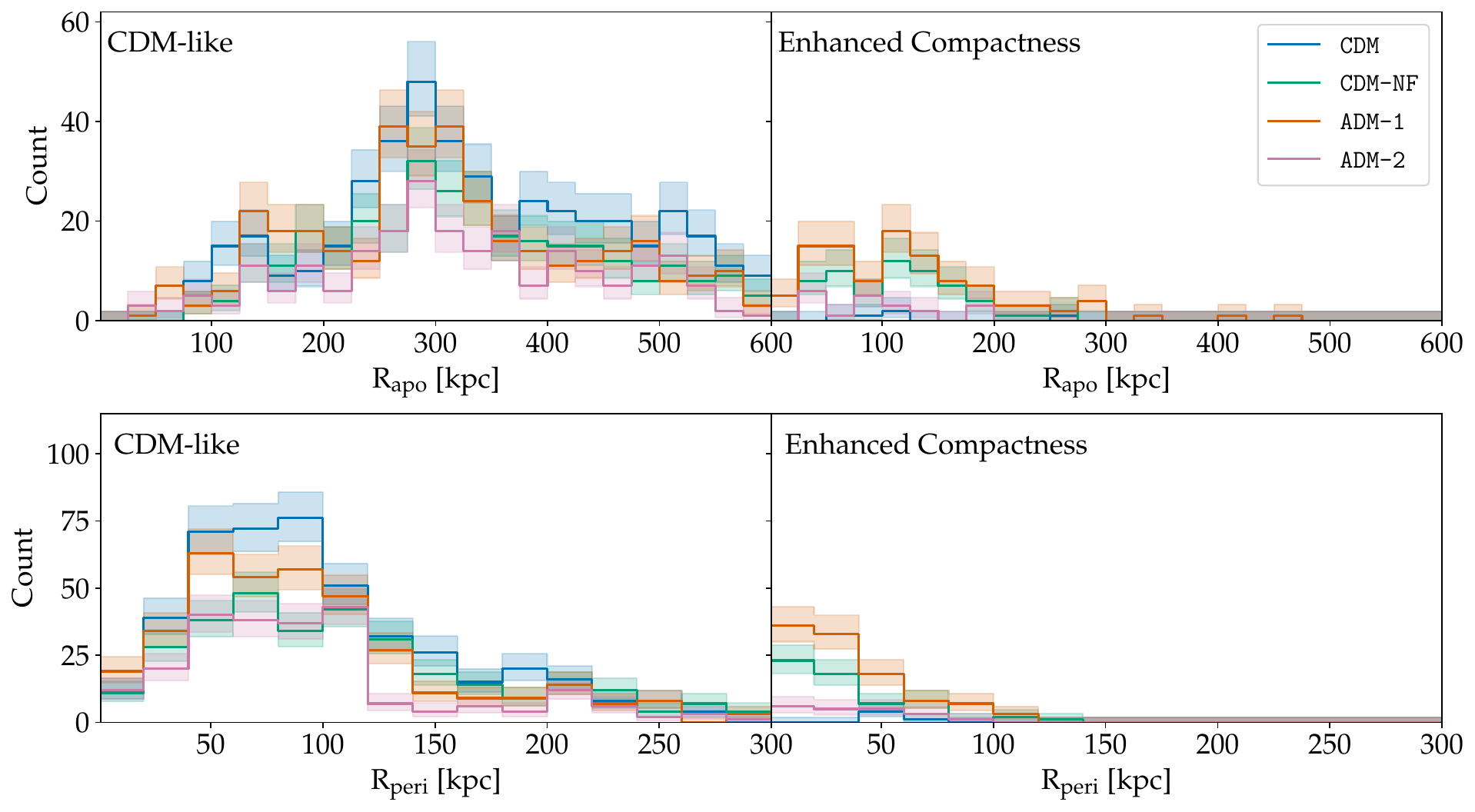}
\caption{Histograms of apocenters and pericenters for DM subhalos. Shaded bands indicate 1$\sigma$ Poissonian errors for each bin. \textbf{Top:} Apocenter distributions for each simulation with CDM-like subhalos ($R_{200, {\rm m}}/R_{1/2}<15$) on the left and Enhanced-compactness subhalos ($15<R_{200, {\rm m}}/R_{1/2}<150$) on the right. The Enhanced-compactness subhalos have much smaller apocenters, occupying orbits predominantly $\lesssim 125$ kpc. \textbf{Bottom:} Pericenter distributions for each simulation with CDM-like subhalos ($R_{200, {\rm m}}/R_{1/2}<15$) on the left  and Enhanced-compactness subhalos ($15<R_{200, {\rm m}}/R_{1/2}<150$) on the right. The Enhanced-compactness subhalos again have a largely different distribution and peak at much smaller values compared to the CDM-like subhalos, indicating these subhalos have undergone close pericentric passages. The enhancement in the lowest bin for pericenters $\lesssim 25$ kpc is not observed in $\cdm$ and supports that this survival of close pericentric passages is possible due to the large central densities found in the \texttt{ADM} subhalos.}
    \label{fig:orbits}
\end{figure*}

To further explore the subhalo radial distributions and this inner enhancement, we now study some of the orbital parameters of the subhalos across simulations. In Fig.~\ref{fig:orbits}, we plot distributions of the subhalo apocenters~(top row) and pericenters~(bottom row). To calculate the apocenters, \texttt{AGAMA}~\citep{Vasiliev_2018} is used to construct an axisymmetric galactic potential from the $z=0$ snapshot of each MW analog using a multipole expansion, which is then used to calculate the apocenter for each constituent particle, with the subhalo's apocenter being the median of these values. The pericenters are determined by tracking the innermost 20 constituent particles at $z=0$ to earlier redshifts, to identify the snapshot at which their median galactocentric distance to the host halo is smallest. To account for the discrete timing of snapshots, \texttt{AGAMA}~\citep{Vasiliev_2018} is again used to construct an axisymmetric galactic potential at this earlier snapshot, from which the constituent particles' pericenters are calculated, with the subhalo's pericenter being the median value.

Considering the distributions in Fig.~\ref{fig:orbits}, the fraction of subhalos with close orbits (R$_{\rm apo} < 125$~kpc) are 0.06, 0.11, 0.16, and 0.11 for $\cdm$, $\cdmnf$, $\admone$, and $\admtwo$, respectively. Of these subhalos, a large fraction are in the Enhanced-compactness group: 0.15~($\cdm$), 0.83~($\cdmnf$), 0.78~($\admone$), and 0.54~($\admtwo$). Thus, there is a larger percentage of subhalos with close orbits (R$_{\rm apo} < 125$~kpc) in the $\cdmnf$ and \texttt{ADM} simulations, mostly due to the population of Enhanced-compactness subhalos. This contributes to the larger numbers of subhalos at small radii found in the \texttt{ADM} and $\cdmnf$ simulations at $z=0$, relative to \texttt{CDM}.

However, the Enhanced-compactness subhalos cannot be the sole cause of the distinctive plateau-like feature near $\sim$125~kpc in the \texttt{ADM} radial distributions, as it would then also appear in $\cdmnf$. Given that it only appears in the \texttt{ADM} simulations, it may be related to the dynamics and formation of the \texttt{ADM} gas disk, which would affect the orbit of the subhalos.  This is supported by the fact the plateau-like feature only becomes apparent at late times ($z \sim 0.2$) in both \texttt{ADM} simulations. We leave a more careful study of the dark disk formation and its impact on the subhalo population to future work. Similarly, while the Enhanced-compactness subhalos in $\cdmnf$ increase the fraction of subhalos with close orbits (R$_{\rm apo} < 125$~kpc), a more careful study of the galactic formation history could comment on the slight relative depletion of total subhalos compared to $\cdm$. 

The pericenter distributions in Fig.~\ref{fig:orbits} show that the Enhanced-compactness distributions are biased to much smaller pericenter values compared to the CDM-like subhalos. In the smallest bin (R$_{\rm peri} < 20$ kpc), the count of Enhanced-compactness~(CDM-like) subhalos for each sim is 0~(12), 23~(11), 36~(19), and 6~(12) for $\cdm$, $\cdmnf$, $\admone$, and $\admtwo$, respectively. In \cite{Garrison_Kimmel_2017}, they found very few subhalos were able to survive to $z=0$ with small pericenters due to the influence of the baryonic disk potential.  The increase in subhalos with small pericenters found in our simulations, predominantly in the Enhanced-compactness group, supports the fact that the concentrated inner densities of the subhalos increase their chance of surviving close pericentric passages.

We now consider the depletion in the total number of subhalos found in $\admtwo$ relative to $\admone$. As previously shown in Fig.~\ref{fig:SHMF}, this is mostly for masses $\lesssim 10^{8.5}\msun$. There are two effects likely contributing to this observed depletion. First is the fact that the $\admone$ subhalos often have larger ADM mass fractions than those in $\admtwo$, which leads to more compact halos that can survive closer orbits, as previously mentioned. The second contributing factor has to do with the formation of a dark disk in the host galaxy. While both \texttt{ADM} simulations form dark gas disks with comparable densities, the $\admone$ disk extends out to radii $\sim3$~kpc, entirely encompassed by the baryonic disk.  Comparatively, the $\admtwo$ disk extends beyond the baryonic disk out to $\sim 14$~kpc. The additional tidal field of a baryonic disk can cause subhalo depletion out to large distances~\citep{Garrison_Kimmel_2017,Silverman:2022bhs}; presumably such an effect would only be enhanced by the presence of a large dark disk.  In light of this, the $\admtwo$ disk is 
much more likely to disrupt the lighter $\admtwo$ subhalos (too light to have been cusped by ADM) that pass nearby, contributing to their overall depletion relative to $\admone$.

To summarize, ADM appears to enhance the \emph{fraction} of subhalos within $\sim125$ kpc of the host, with ADM cooling in subhalos creating cuspier profiles that increase their likelihood of surviving small pericentric passages. However, it is also possible that the dynamics of the \texttt{ADM} disk formation at late times also causes an enhancement. Lastly, this can result in either an enhancement or reduction in the total number of subhalos, depending on the extent of the ADM gas disk that forms and the number / mass of subhalos containing ADM clumps. The potential enhancement in the number of subhalos with apocenters within 125~kpc of the galactic center also increases the number of subhalos within the range of stellar streams ($R_{\rm host} \lesssim 50$~kpc), making ADM an interesting candidate for future stellar stream surveys. Furthermore, since the subhalos in this region tend to be more compact than the CDM expectation, the induced perturbations on the stellar structure will be predictably different~\citep{Bonaca:2018fek}.

\subsection{Subhalos in the Disk Plane}
\label{sec:clumps}

In this subsection, we briefly discuss the population of subhalos that are found in the baryonic disk and its immediate vicinity. These subhalos are primarily comprised of the ADM-dominated population (see Fig.~\ref{fig:rhalf}) with $R_{200, {\rm m}}/R_{1/2} \gtrsim 150$ and consist almost entirely of ADM clumps. 
Because ADM clumps are an order-of-magnitude lighter than CDM particles, these subhalos can be resolved down to $\sim 10^6 \msun$.

\begin{figure*}
    \centering
    \includegraphics[width=0.49\linewidth]{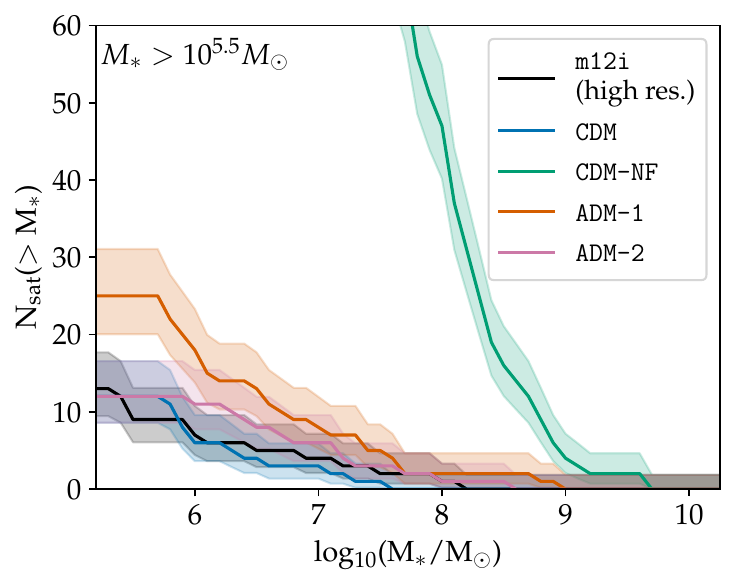}
    \includegraphics[width=0.49\linewidth]{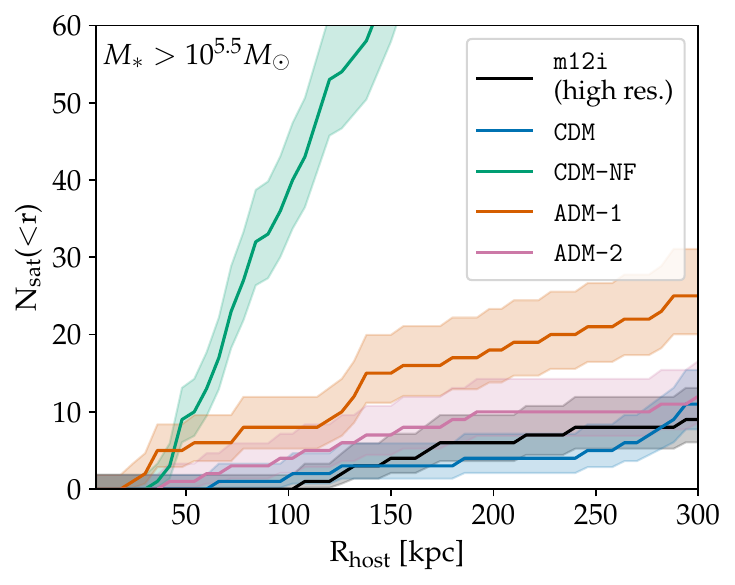}
    \caption{Properties of visible satellite galaxies ($M_* \gtrsim 10^{5.5} \msun$) within 300~kpc of their host at $z=0$. Shaded bands indicate the 68\% containment in the count. The black solid line depicts the high-resolution~(high res.) \texttt{FIRE-2 m12i} simulation. \textbf{Left:} Cumulative number of galaxies above a given stellar mass. $\cdmnf$ evidently over-produces satellite galaxies, as well as making them too massive. For this reason, the $\cdmnf$ simulation will be suppressed for the rest of the paper. $\admone$ also has a small enhancement in total number, specifically in the lower stellar mass range. $\admtwo$, $\cdm$, and the high-resolution \texttt{m12i} all show good agreement within the observed spread. \textbf{Right:} Cumulative number of satellite galaxies within a given galactocentric radius. The radial distribution for $\admtwo$, $\cdm$, and the high-resolution \texttt{m12i} all exhibit similar trends, while $\admone$ differs in having a larger number of visible satellite galaxies across all host distances. $\cdmnf$ vastly overproduces satellite galaxies both in stellar mass and number.}
    \label{fig:stellar_distributions}
\end{figure*}

The ADM clumps in these subhalos all formed at late times $(z \lesssim 2)$ and within 10~kpc of the center of the galaxy. To characterize their orbits, we use the orbital circularity, $\epsilon = j_z/j_c(E)$, defined as the ratio of the $z$-component of a particle's angular momentum to the angular momentum of a particle with a circular orbit and the same energy \citep{Abadi:2002tt}. Of the ADM-dominated subhalos, 80\% have thin-disk orbits ($\epsilon > 0.8$) and lie at heights $z < 2$~kpc from the disk plane. Extending the circularity range to thick-disk orbits ($\epsilon > 0.2$), 93\% of these subhalos are now included.  This strongly suggests that the ADM-dominated subhalos form near the disk plane, likely from fragmentation of the ADM disk in the host's central region. Because $\admtwo$ has a larger cut-off temperature due to its higher binding energy, its dark disk is more resistant to fragmentation, which explains why it forms fewer ADM-dominated subhalos.

Because these subhalos lie within the disk region of MW-like galaxies and consist primarily of ADM clumps, which are dark compact objects at our simulations' resolution level, their abundance could be constrained from gravitational microlensing events~\citep{Winch_2022}, in addition to the microlensing produced by the individual dark stars or ADM-sourced black holes. 
Microlensing from extended structures has been considered and could be applied to these compact subhalos~\citep{Croon:2020wpr}. However, at the current resolution, these disk subhalos have an average half-mass radius of 13~pc, much larger than the maximum Einstein radius expected for surveys of stars in the bulge (e.g., 0.02 pc), suggesting that these subhalos will provide weak to no constraints when acting as lenses. 

Additional constraints on these compact objects could come from interactions with wide binaries \citep{1985ApJ...290...15B,2014ApJ...790..159M}, where fly-by encounters with the dark compact objects inject energy into the two-body system, increasing its semi-major axis or even completely disrupting the system. Recent work has also used these observations to provide some constraints on extended DM subhalos \citep{Ramirez:2022mys}, up to $\mathcal{O}(1$ pc) scales. Again, probing such phenomena is beyond the resolution of our current simulations, 
but this demonstrates that ADM-dominated subhalos in the disk plane could lead to distinct observables and emphasizes the importance of incorporating future improvements to ADM  simulations, such as dark molecular cooling~\citep{Ryan:2021tgw} that will allow one to resolve ADM clumps to smaller scales. 

Lastly, compact objects can also heat the interstellar medium (ISM) via dynamical friction~\citep{Carr_1999}, providing constraints on their abundance. The effect of compact subhalos on the ISM has also been considered in \cite{Wadekar:2022ymq}, who showed that observations of the gas-rich dwarf galaxy Leo T can constrain the fraction of DM in subhalos with properties similar to ADM-dominated subhalos. Future work studying the subhalo distribution of ADM in dwarf galaxies could thus potentially yield constraints using this methodology.

\section{Visible Satellite Galaxies}
\label{sec:sat galaxies}

This section explores the properties of the subset of subhalos with stellar mass comparable to that of the MW's dwarf galaxies.  To this end, we now require that each subhalo contain at least 10 stellar particles, which is equivalent to a stellar mass cut of approximately $M_*\gtrsim 10^{5.5} \msun$, as outlined in Sec.~\ref{sec:analysis}. This respectively selects a total of 12, 105, 25, and 12 subhalos from $\cdm$, $\cdmnf$, $\admone$, and $\admtwo$.

Figure~\ref{fig:stellar_distributions} shows both the cumulative stellar mass~(left panel) and radial distribution~(right panel) of visible satellite galaxies across the simulations. There is good agreement (to within the observed spread) between $\cdm$ and the high-resolution \texttt{m12i} for both mass and radial distributions. This supports that the \texttt{CDM} simulation has converged for this range of satellite galaxy masses. Unsurprisingly, the $\cdmnf$ simulation overproduces satellite galaxies both in stellar mass and number. We suppress showing $\cdmnf$ in the remaining figures for clarity, as it clearly diverges from a realistic MW satellite population.

Comparing stellar masses, $\log_{10}(M_*/\msun)$, of the satellite galaxies, the median and 68\% containment values for $\admone$~($6.44_{-0.71}^{+0.99}$), $\admtwo$~($6.88_{-0.65}^{+0.74}$) and $\cdm$~($6.05_{-0.30}^{+0.97}$) are all consistent  within the observed spread, the only difference being that $\admone$ appears to overproduce specifically lighter satellites with M$_*<10^{6}\msun$. In general, satellite galaxy formation is enhanced in $\texttt{ADM}$ simulations.  For example, $\admone$ has roughly twice as many visible satellites as $\cdm$, even though it only produces 20\% more subhalos overall. $\admtwo$ forms 60\% of the number of subhalos compared to $\cdm$, but still has a comparable number of visible satellites.
The visible satellites in both simulations are part of the Enhanced-compactness population (see Fig.~\ref{fig:rhalf}), with the majority of visible satellites in $\admone$~(84\%) and $\admtwo$~(100\%) having $R_{200,\rm{m}}/R_{1/2} \gtrsim 15$. Comparatively, the maximum compactness for visible $\cdm$ satellites is 11. This is an important point to emphasize: Because ADM increases the central potentials of subhalos, the ADM-containing subhalos are more likely to trap Standard Model baryons and form visible satellite galaxies. For the case of $\admtwo$, only 16\% of subhalos contain ADM, but it is only this subset that go on to host satellite galaxies.

\begin{figure}
\centering
\includegraphics[width=\linewidth]{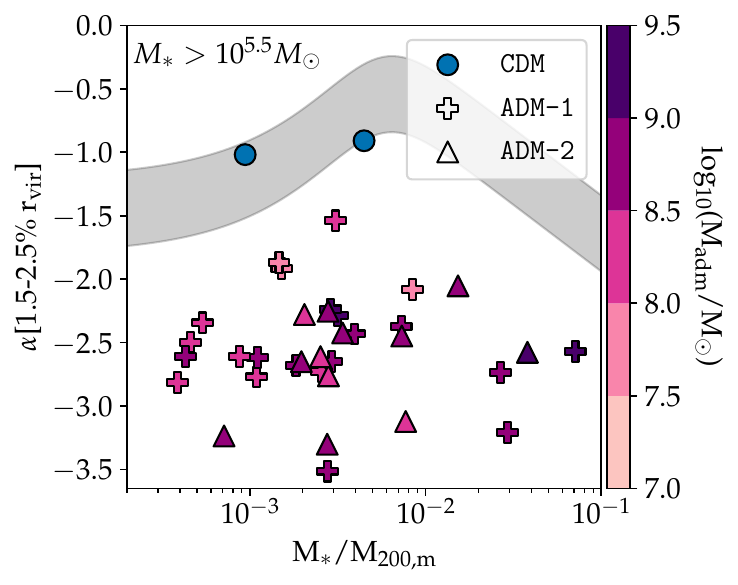}
\caption{Inner DM density slope, $\alpha$, averaged over $(1.5$--$2.5\%)\times \rm{r}_{\rm vir}$, as a function of the stellar mass fraction, $M_* / M_{200,\rm{m}}$, at $z = 0$. The \texttt{ADM} points are shaded to indicate the total ADM clump mass contained in each subhalo. The gray band shows the density slope from FIRE-2 CDM simulations of field galaxies, taken from \cite{Lazar_2020}. While only a few $\cdm$ satellite galaxies are resolved at these inner radii, most \texttt{ADM} satellites are. The inner regions ($r<0.5$~kpc) of these \texttt{ADM} satellites are dominated by the collapsed ADM clumps and form extremely cuspy slopes, with median values $\alpha\sim-2.6$.} 
\label{fig:slope}
\end{figure}

To quantify this enhancement in the central density profile, we plot the slope of the density profile, $\alpha$, at small radii. Specifically, we fit the density to a power law, $\rho (r) \propto r^\alpha$, for radii in the range $(1.5$--$2.5\%)\times \rm{r}_{\rm vir}$, where $r_{\rm vir}$ is the virial radius.  Only satellite galaxies with $r_{\rm{DM}}^{\rm{conv}} < 1.5\% \times $r$_{\rm vir}$ are included. Note that this excludes the majority of $\cdm$ satellite galaxies. The results for the $\cdm$ subhalos are shown by the blue circles in Fig.~\ref{fig:slope}. For comparison, the gray band shows the best-fit model for the FIRE-2 feedback physics, as obtained in~\cite{Lazar_2020}.\footnote{Note that~\cite{Lazar_2020} uses a smaller radial range $(1$--$2\%)\times \rm{r}_{\rm vir}$. We increase the range to ensure a subset of the $\cdm$ satellites have converged at $1.5\%$ r$_{\rm vir}$, but we find that it makes little difference for the \texttt{ADM} slopes in this range---both their median values decrease slightly from $\alpha = -2.7$ to $-2.6$ when increasing the range. Additionally, ~\cite{Lazar_2020} uses field galaxies to determine the depicted trend, while our simulations are all satellite galaxies, thus experiencing additional tidal forces from the host galaxy. The observed differences between $\cdm$ and both $\texttt{ADM}$ simulations are robust.} The overall trend is similar to the distributions for MaGICC~\citep{DiCintio:2013qxa} and NIHAO~\citep{Tollet_2016}, although the slope distributions are not the same in detail.

\begin{figure}
    \centering
    \includegraphics[width=\linewidth]{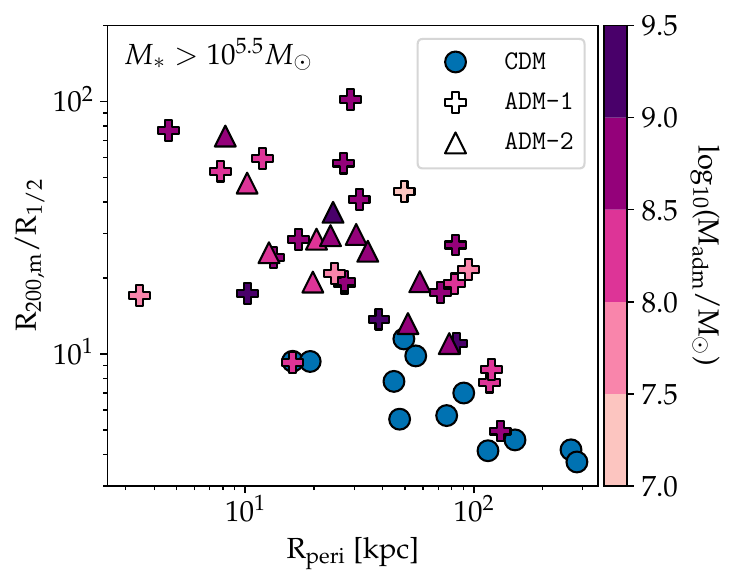}
    \caption{Total DM compactness, R$_{200,\rm{m}}/$R$_{1/2}$, as a function of a visible satellite galaxy's orbital pericenter, $R_{\rm peri}$. The \texttt{ADM} data points are shaded to indicate the total ADM clump mass contained in each subhalo. The results show a trend of increasing compactness as a subhalo's pericenter decreases. Specifically, a large fraction of \texttt{ADM} subhalos reach smaller pericenters compared to their $\cdm$ counterparts.} 
    \label{fig:pericenter}
\end{figure}
The $\cdm$ satellites lie on the band from \cite{Lazar_2020} and have slopes with $\alpha\sim-1$, but we cannot make any strong statements with only two data points. The visible satellites in the \texttt{ADM} simulations largely differ from the CDM expectation. The profiles are cuspier than NFW, with median $\alpha$ values of $-2.6^{+0.4}_{-0.2}$ for $\admone$ and $-2.6^{+0.3}_{-0.6}$ for $\admtwo$.  (The lower/upper ranges denote 68\% containment.) This is because the ADM clumps mainly occupy a subhalo's central-most region, with an average of 69\% contained within $2.5\%\times r_{\rm vir}$ across all \texttt{ADM} satellite galaxies. Thus, for the parameter points considered in these $\texttt{ADM}$ simulations, visible satellites are typically all produced with significant inner cusps, rather than with a diversity of inner slopes.

The spatial distribution of visible satellites is a consequence of the individual orbits of its members.  A relevant relation to consider is thus the pericentric distances for the visible satellites and how they correlate with compactness. The pericenters are calculated using the same methodology described in Sec.~\ref{sec:orbits} using the stellar particles contained in the satellite galaxy.

Figure~\ref{fig:pericenter} shows the relationship between a visible satellite's pericenter and its compactness.
As mentioned earlier, the visible satellites in $\admone$ and $\admtwo$ are typically more compact than those in $\cdm$, but they also have smaller pericenters, as shown by the median and 68\% containment values for $\admone$ ($27.2^{+58.3}_{-17.4}$~kpc) and $\admtwo$ ($23.9^{+29.2}_{-11.8}$~kpc) compared to $\cdm$
($65.9^{+113.2}_{-27.3}$~kpc). This pattern is suggestive of a survivorship bias, where satellites with large central densities are those that survive passages with small pericenters. As mentioned previously, cuspy profiles are more resistant to tidal stripping than cored profiles~ \citep{Errani_2016}, which supports this interpretation.

\begin{figure}
    \centering
    \includegraphics[width=\linewidth]{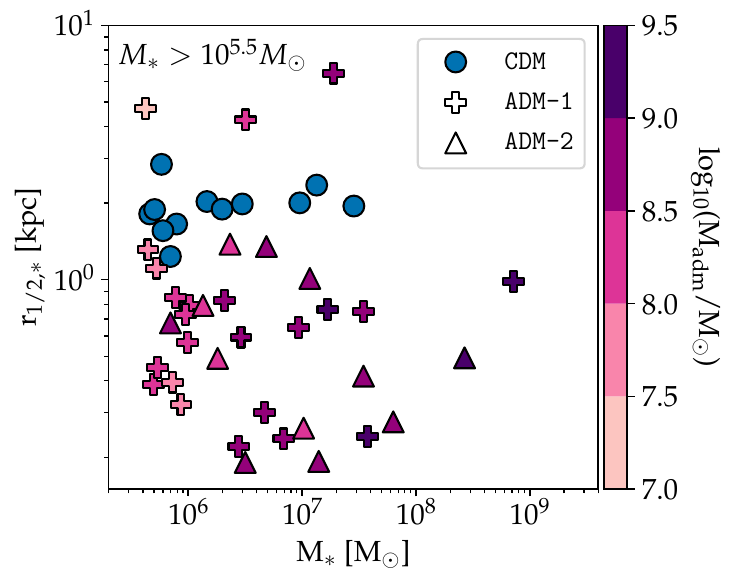}
    \caption{Stellar-half-mass radius, $r_{1/2,*}$, versus stellar mass, $M_*$, for the visible satellite galaxies in each simulation, with the \texttt{ADM} points  shaded to indicate the total ADM clump mass contained in each subhalo. On average, both \texttt{ADM} simulations form more compact satellites compared to the $\cdm$ simulation, but with a larger distribution of radii.} 
    \label{fig:stellar_radii}
\end{figure}

We now inspect whether the increased compactness of the DM subhalos in $\admone$ and $\admtwo$ affects the spatial distribution of their stars. Figure~\ref{fig:stellar_radii} plots the relationship of stellar-half-mass radius, $r_{1/2,*}$, the radius that encloses 50\% of the galaxy's stellar mass, with $M_*$. We require that  $r_{1/2,*}$ be larger than $r^{\rm{conv}}_{\rm{DM}}$ for each satellite galaxy to be confident the DM distribution has converged.  The baryons, however, are susceptible to further resolution effects such as an artificially enhanced `bursty' star formation rate \citep{Hopkins:2017ycn}, which can puff up satellite galaxies. These effects are discussed in Appendix~\ref{sec:resolution}, where we compare to lower-resolution runs to check the validity of our results and find that the qualitative trends are robust, even if the quantitative values may shift.  In general, the \texttt{ADM} galaxies have more compact stellar distributions, with smaller $r_{1/2, *}$ values for $\admone$ ($0.75^{+3.57}_{-0.43}$~kpc) and $\admtwo$ ($0.49^{+0.60}_{-0.25}$~kpc) compared to $\cdm$ ($1.92^{+0.19}_{-0.29}$~kpc). The range in the $\admone$ values has a small overlap  with $\cdm$ as it produces some puffier outlier satellite galaxies. These satellites are also interesting to highlight as they indicate that while $\admone$ is routinely producing subhalos with very cuspy inner slopes, it is still possible to produce a  diversity of baryonic structure, such as these  diffuse galaxies with relatively large $r_{1/2, *}$ values.

\begin{figure}
    \centering
    \includegraphics[width=\linewidth]{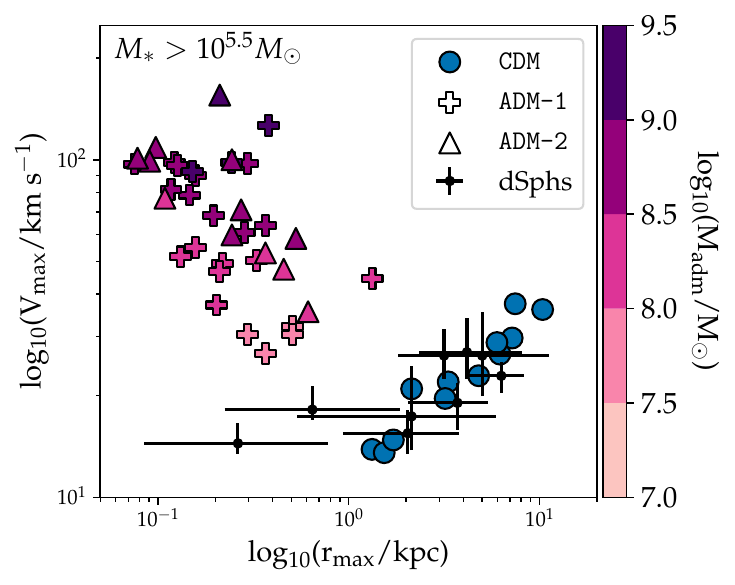}
    \caption{V$_{\rm max}$ versus r$_{\rm max}$ for the visible satellite galaxies. The values correspond to the maximum velocity of the rotation curve and the radial position of this peak. The black points show the corresponding data for the Milky Way dwarf spheroidals, as obtained in \cite{Kaplinghat:2019svz}, assuming an NFW DM profile. The \texttt{ADM} points are shaded to indicate the total ADM clump mass contained in each subhalo, which is strongly correlated with the resulting V$_{\rm{max}}$ for the \texttt{ADM} satellite galaxies. We observe a clear distinction between the \texttt{ADM} and $\cdm$ satellites, which agree relatively well with the observed Milky Way dwarf spheroidals.} 
    \label{fig:rmax}
\end{figure}

All of this informs what may be the most distinguishing, and perhaps most robust, characteristics displayed by visible satellites with ADM compared to CDM-only. Figure~\ref{fig:rmax} shows the (r$_{\rm max}$,V$_{\rm max}$) values for the satellite galaxies across simulations, where (r$_\mathrm{max}, $V$_\mathrm{max}$) defines the maximum of its circular velocity curve. Again, we ensure that r$_{\rm{max}}>r^{\rm{conv}}_{\rm{DM}}$ for each satellite galaxy, which results in one $\admone$ satellite galaxy being excluded. Because the subhalos in $\admone$ and $\admtwo$ have increased central densities, their V$_{\rm max}$ values are enhanced compared to those from the $\cdm$ simulation and occur at smaller radii.  The (r$_{\rm max}, $V$_{\rm max}$) values for $\cdm$ correspond to  $(3.69_{-2.04}^{+3.22}~\text{kpc}, 22.44_{-7.88}^{+8.74}~\text{km/s})$, while the $\admone$ and $\admtwo$ simulations have  
($0.22_{-0.09}^{+0.16}$~kpc, $62.32_{-26.83}^{+34.84}$~km/s) and
($0.24_{-0.15}^{+0.23}$~kpc, $74.00_{-22.35}^{+28.93}$~km/s). This leads to a large separation between $\cdm$ and both $\texttt{ADM}$ simulations in the r$_{\rm max} - $V$_{\rm max}$ plane.  By eye, the two \texttt{ADM} simulations do not produce a population of CDM-like satellite galaxies. There also appears to be a correlation with V$_{\rm max}$ and the total ADM mass enclosed within the subhalo, depicted by the shading of \texttt{ADM} parameter points. This  correlation is seemingly stronger than the dependence on the specific cooling physics, as $\admone$ and $\admtwo$ have similar distributions. 

Overlaid on the plot are r$_{\rm max} - $V$_{\rm max}$ values for bright dwarf spheroidals~(dSph), originally presented in \cite{Kaplinghat:2019svz}. In their work, they use Jeans analysis to infer projected line-of-sight velocities and perform a Bayesian fit to the observed dSph data assuming an NFW profile for the DM, which is in good agreement with the $\cdm$ simulation. Because an NFW profile is assumed when performing the Bayesian fit to the data, and the \texttt{ADM} subhalos favor a cuspier density profile, the data points are best compared to the $\cdm$ simulation. To directly compare the \texttt{ADM} simulation to data, one would have to implement a Bayesian fit assuming some parameterised ADM density profile.

As discussed in Appendix~\ref{sec:resolution}, the qualitative discrepancy between \texttt{ADM} and $\cdm$ satellites is quite resolution robust. Resolution-dependent effects tend to puff up satellite galaxies, counteracting the density enhancement provided by the ADM. Therefore any qualitative changes higher-resolution runs might cause will only increase this separation between the \texttt{ADM} and $\cdm$ satellites. We can therefore claim with some confidence that both ADM simulations only form highly compact satellite galaxies that do not resemble the satellite population we observe in the MW.

\section{Conclusions}
\label{sec:conclusion}

Compared to CDM simulations, we find that our two simulations with a 6\% ADM subcomponent cause large deviations in the properties of DM substructure, demonstrating that even small ADM fractions can significantly affect galactic dynamics.
The various differences can all be traced back to the fact that the ADM gas is able to cool and collapse in the center of subhalos, leading to enhanced central densities and more compact halos overall. This is realized by the ADM gas itself cooling and collapsing to clumps in the central region, as well as the gas dragging the CDM inward via dark baryonic contraction. This increased central density makes subhalos more resistant to tidal disruption, allowing them to survive small pericentric passages and occupy orbits at smaller radii, which is potentially causing an increase in the fraction of subhalos in the inner regions of the host galaxy.

The increased subhalo compactness affects a range of properties for visible satellite galaxies in the \texttt{ADM} simulations. Compared to satellites in CDM simulations, they have cuspier inner slopes and smaller stellar-half-mass radii.
Thus, ADM is able to produce more compact satellites due to its efficient cooling, see Fig.~\ref{fig:slope}, but these satellites fail to produce the full diversity of inner slopes observed in data.
In fact, for the parameter points in our simulations, satellites with ADM were completely distinct from the CDM-only expectation in the $(r_\mathrm{max}, V_\mathrm{max})$-plane, see Fig.~\ref{fig:rmax}.
This robustly extends the conclusions of \cite{Roy:2023zar}, in particular the exclusion of $\admone$, which does not disrupt the Standard Model gas disk in the central host galaxy as obviously as $\admtwo$.

Additionally, we highlighted the ability of ADM to form DM subhalos not through cosmic overdensities but through fragmentation of the ADM gas disk. While future work is needed, these highly concentrated dark objects could prove to be a promising supplementary source of constraints from microlensing and wide binary surveys.

These results motivate a more extensive systematic investigation of dwarf galaxy properties in the presence of an ADM sub-component. 
This will determine the ranges of parameters for which ADM is excluded due to its impact on galactic dynamics, as well as the ADM parameters that are compatible with observations.
Such a study would place new and powerful constraints on the ADM parameter space, likely with far greater sensitivity than purely cosmological probes~\citep{Bansal:2022qbi}.
An even more exciting possibility would be the discovery of ADM using current or future observation data of satellite galaxies, aided by the detailed understanding of its effects that these future studies would provide.
This may also settle the question of whether the more compact satellite galaxies produced by an ADM subcomponent could ultimately address the diversity problem.

While we are only at the start of assembling a detailed and general understanding of ADM in our Galaxy,
the initial studies already demonstrate that dissipative DM interactions can have significant impacts on galactic-scale physics.  The implications of these findings extend beyond this specific case study of ADM, as they demonstrate that non-trivial dark sectors leave stark imprints on the formation and dynamics of galaxies.  This motivates deeper study of a broader range of dark-sector scenarios along with the corresponding astrophysical tests needed to successfully probe them.  

\section{Acknowledgments}
We would like to thank Arpit Arora, Ting Li, Tri Nguyen, Xiaowei Ou, Nondh Panithanpaisal, Robyn Sanderson, and Nora Shipp for helpful discussions during the completion of this project. The research of DC and CG was supported in part by Discovery Grants from the Natural Sciences and Engineering Research Council of Canada and the Canada Research Chair program. The research of DC was also supported by the Alfred P. Sloan Foundation, the Ontario Early Researcher Award, and the University of Toronto McLean Award. The work of CG was also supported by the University of Toronto Connaught International Scholarship, McDonald Institute Graduate Student Exchange program, and the Canada First Research Excellence Fund. ML is supported by the Department of Energy~(DOE) under Award Number DE-SC0007968, as well as the Simons Investigator in Physics Award. This work was performed in part at the Aspen Center for Physics, which is supported by National Science Foundation grant PHY-2210452. The work presented in this paper was performed on computational resources managed and supported by Princeton Research Computing. This research also made extensive use of the publicly available codes \texttt{IPython}~\citep{PER-GRA:2007}, \texttt{matplotlib}~\citep{Hunter:2007},  \texttt{Rockstar} \citep{Behroozi_2012},
\texttt{Jupyter}~\citep{Kluyver2016jupyter},
\texttt{AGAMA}~\citep{Vasiliev_2018},
\texttt{HaloAnalysis}~\citep{2020ascl.soft02014W},
\texttt{NumPy}~\citep{harris2020array}, 
\texttt{SciPy}~\citep{2020SciPy-NMeth}, and 
\texttt{gizmo-analysis}~\citep{Wetzel2020}.

\bibliography{adm_subhalos}

\appendix

This appendix discusses several systematic uncertainties relevant for the analysis in the main body and their impact on the primary conclusions of the work.  In particular, the choice of particle resolution, friends-of-friends linking percentage, and subhalo threshold criteria are discussed.\vspace{0.2in}

\section{Resolution Effects on Visible Satellites}
\label{sec:resolution}

\setcounter{equation}{0}
\setcounter{figure}{0} 
\setcounter{table}{0}
\renewcommand{\theequation}{A\arabic{equation}}
\renewcommand{\thefigure}{A\arabic{figure}}
\renewcommand{\thetable}{A\arabic{table}}

The properties of low-mass satellite galaxies can be resolution-dependent, with the dark matter~(DM) and stellar content being `puffed up' to larger sizes~\citep{Shen:2022opd}. For example,~\cite{Fitts_2019} found that increasing the baryon mass resolution from $4000 \msun$ to $62.5 \msun$ resulted in their r$_{1/2,*}$ values shrinking by half in a study of the isolated dwarf, \texttt{m10b}. \cite{10.1093/mnras/stad2615} also found that lower-resolution simulations could cause DM particles to spuriously heat stellar matter, leading to satellite galaxies with larger sizes and characteristic velocities when compared to higher-resolution simulations.

Here, we explore how the main-text results concerning visible satellite galaxies depend on resolution.  In particular, we compare to results from a $\cdm$, $\admone$, and $\admtwo$ simulation that are run at \emph{lower} resolution to the primary simulations in the paper. The relevant particle masses for the low-resolution simulations are provided in Tab.~\ref{tab:ADMspecies_lowres}. 

 \begin{table*}[h!]
\footnotesize
\begin{center}
\renewcommand{\arraystretch}{1.5}
\begin{tabular}{c|c|cccccccccc}
  \Xhline{3\arrayrulewidth}
\textbf{Simulation}&\textbf{Included Species} &$\mathbf{\frac{\Omega_{\rm cdm}}{\Omega_{\rm m}}}$ & $\mathbf{\frac{{m_{\rm cdm}}}{{\msun}}}$ & $\mathbf{\frac{\Omega_{\rm adm}}{\Omega_{\rm dm}}}$ &  $\mathbf{\frac{m_{\rm adm}}{\msun}}$ & $\mathbf{\frac{\Omega_{\rm b}}{\Omega_{\rm m}}}$ & $\mathbf{\frac{m_{\rm b}}{\msun}}$ & $\boldsymbol{\frac{\alpha'}{\alpha}}$ & $\mathbf{\frac{m_{p'}}{m_p}}$& $\mathbf{\frac{m_{e'}}{m_e}}$ & $\mathbf{\frac{T_{\rm cmb'}}{T_{\rm cmb}}}$\\
\hline
$\cdm$ \texttt{(LOW RES)} & CDM+Bar. & 0.83 & 2.25$\times 10^{6}$ & 0 & - & 0.17 & $4.0\times10^5$ & - & - & - & -   \\
$\admone$ \texttt{(LOW RES)} & CDM+ADM-1+Bar.  & 0.78 & 2.12$\times 10^{6}$ & 0.06 &  1.35$\times 10^{5}$ & 0.17 & $3.9\times10^5$ &$1/\sqrt{0.55}$ & $1.3$ & $0.55$ & 0.39  \\
$\admtwo$ \texttt{(LOW RES)} & CDM+ADM-2+Bar.  & 0.78 & 2.12$\times 10^{6}$ & 0.06 &  1.35$\times 10^{5}$ & 0.17 & $3.9\times10^5$ &$2.5$ & $1.3$ & $0.55$ & 0.39  \\
  \Xhline{3\arrayrulewidth}
\end{tabular}
\end{center}
\caption{\label{tab:ADMspecies_lowres} Similar to Tab.~\ref{tab:ADMspecies}, except listing the properties for the low-resolution versions of each simulation.}
\end{table*}

\begin{figure*}[h!]
    \centering
    \includegraphics[width=0.49\linewidth]{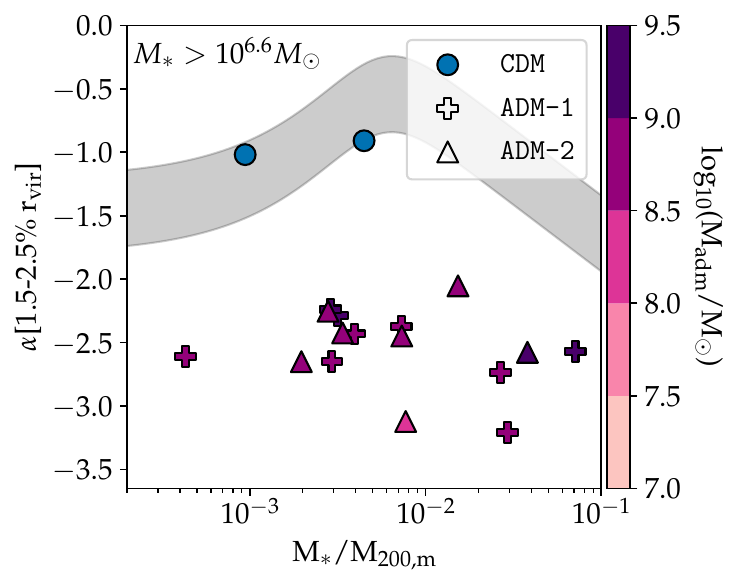}
    \includegraphics[width=0.49\linewidth]{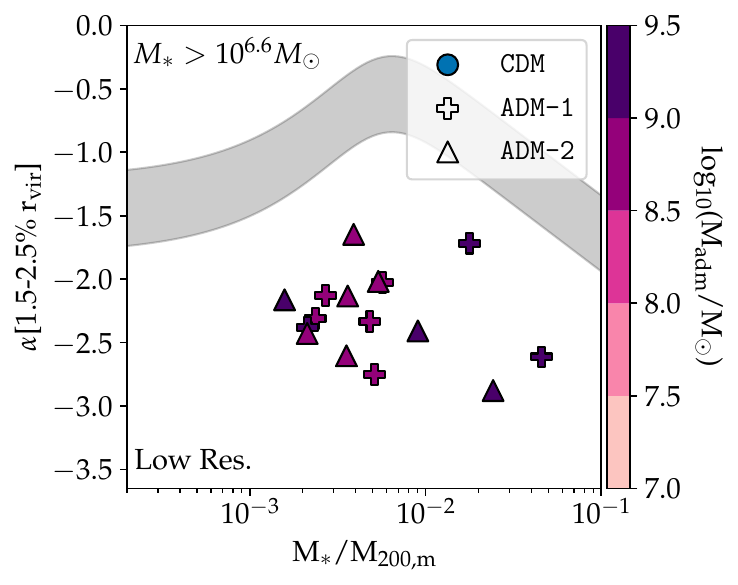}
    \caption{The left panel uses the same simulations as Fig.~\ref{fig:slope}, but with a larger stellar mass cut, $M_* > 10^{6.6} \msun$, and the right panel is its lower-resolution counterpart with the same stellar mass cut. All subhalos plotted satisfy the convergence criterion, $r_{\rm{DM}}^{\rm{conv}} < 1.5\% \times $r$_{\rm vir}$, excluding more of the lower-resolution satellite galaxies. The resolved \texttt{ADM} satellites have a similar distribution of inner slopes, agreeing within the observed scatter. } 
    \label{fig:res_slope}
\end{figure*}

Figure~\ref{fig:res_slope} shows the relationship between the stellar mass fraction ($M_*/M_{200,\rm{m}}$) and the inner slope of the DM density profile. The left panel shows the same simulations as Fig.~\ref{fig:slope}, while the right panel shows the corresponding result for the lower-resolution simulations. The increased baryon particle mass in the lower-resolution simulation results in a larger stellar mass cut as we require each satellite galaxy to contain at least 10 baryon particles in the lower-resolution simulations. We implement a larger stellar mass cut, $M_* > 10^{6.6} \msun$, for both panels to directly compare the results. Again, we require all subhalos to satisfy the convergence criterion, $r_{\rm{DM}}^{\rm{conv}} < 1.5\% \times $r$_{\rm vir}$, which excludes all $\cdm$ satellite galaxies in the lower-resolution simulation. With the increased stellar mass cut, the equivalent values for the higher-resolution simulations are  $-2.59_{-0.41}^{+0.26}$~($\admone$) and  $-2.45_{-0.22}^{+0.20}$~($\admtwo$). In the lower-resolution simulations, there is still a population of resolved \texttt{ADM} satellites with median slopes and 68\% containment values of  $-2.32_{-0.27}^{+0.28}$~($\admone$) and  $-2.29_{-0.30}^{+0.25}$~($\admtwo$).  The median values for the pairs of simulations agree within the observed scatter even though the qualitative values do shift. Overall the observation that the \texttt{ADM} satellites form cuspier density profiles to the trends expected from CDM with baryonic feedback is robust. 

\begin{figure*}[h!]
    \centering
    \includegraphics[width=0.49\linewidth]{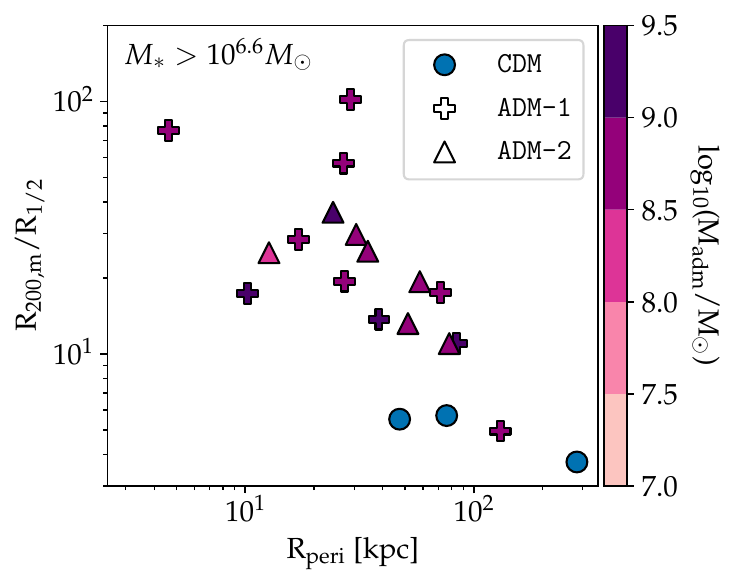}
    \includegraphics[width=0.49\linewidth]{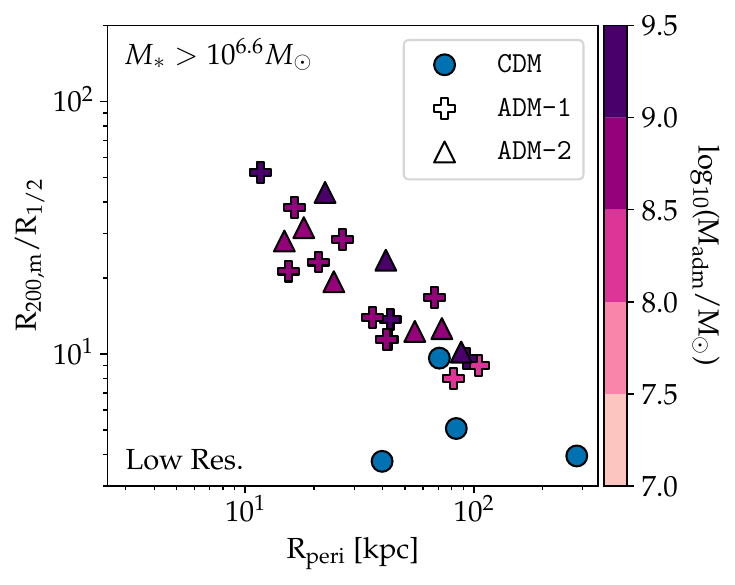}
    \caption{The left panel uses the same simulations as Fig.~\ref{fig:pericenter}, but with a larger stellar mass cut, $M_* > 10^{6.6} \msun$, and the right panel is its lower-resolution counterpart with the same stellar mass cut. We find the same trend in both plots, with a general correlation between a satellite galaxy's compactness and its pericenter.} 
    \label{fig:res_pericenter}
\end{figure*}

Figure~\ref{fig:res_pericenter} shows the relationship between pericenter and compactness for each satellite galaxy. The left panel shows the same simulations as Fig.~\ref{fig:slope}, but with an increased stellar mass cut ($M_* > 10^{6.6} \msun$), while the right panel shows the corresponding result for the lower-resolution simulations. We see the same trends, even more evident in the lower-resolution \texttt{ADM} simulations, of a correlation between a satellite galaxy's compactness and pericenter. The higher-resolution simulations with the new mass cut have pericenters $76.1_{-19.5}^{+140.5}$~kpc~($\cdm$), $28.1_{-14.8}^{+50.3}$~kpc~($\admone$), and
$34.4_{-10.6}^{+24.5}$~kpc~($\admtwo$) and compactness $5.52_{-1.21}^{+0.13}$~($\cdm$), $18.5_{-6.26}^{+49.6}$~($\admone$), and
$25.2_{-12.1}^{+4.84}$~($\admtwo$). The lower-resolution simulations have median pericenters with 68\% containment  of $77.3_{-22.7}^{+110}$~kpc ($\cdm$), $41.4_{-25.0}^{+41.2}$~kpc~($\admone$), and
$32.8_{-14.3}^{+37.6}$~kpc~($\admtwo$) and  compactness $4.51_{-0.66}^{+2.93}$~($\cdm$), $14.0_{-4.43}^{+15.3}$~($\admone$), and
$21.4_{-9.09}^{+9.78}$~($\admtwo$). Again, these median values agree within the observed scatter.

\begin{figure*}[h!]
    \centering
    \includegraphics[width=0.49\linewidth]{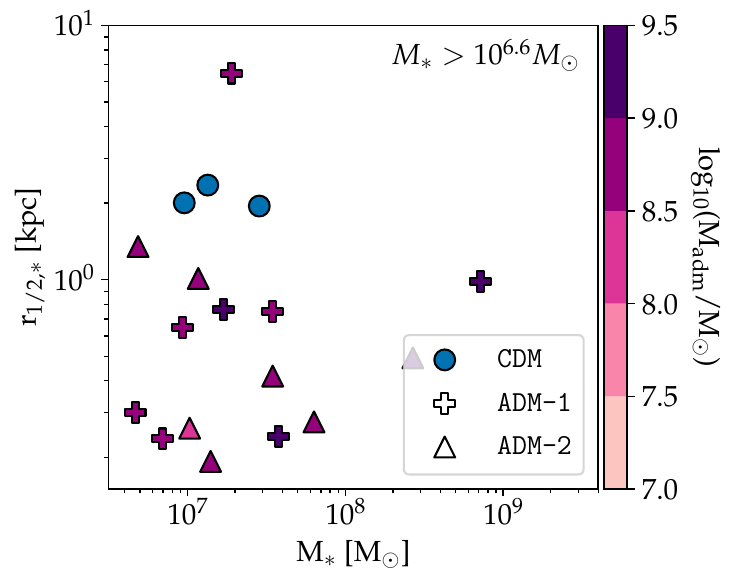}
    \includegraphics[width=0.49\linewidth]{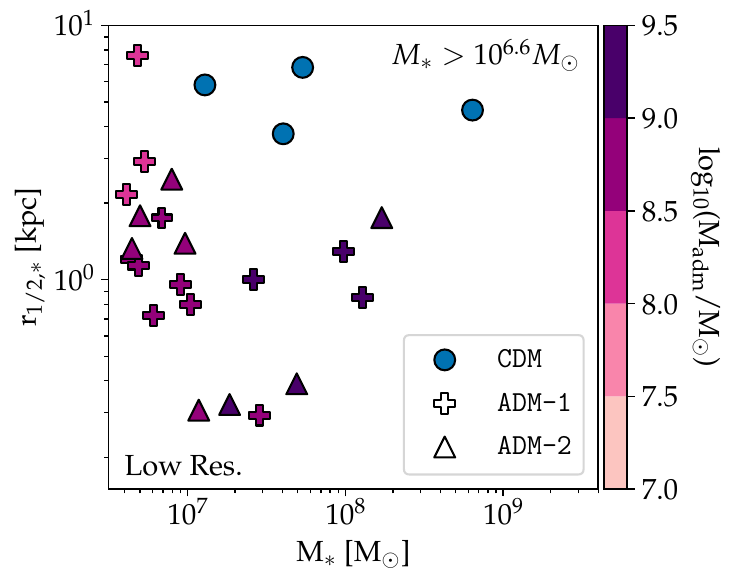}
    \caption{The left panel uses the same simulations as Fig.~\ref{fig:stellar_radii}, but with a larger stellar mass cut, $M_* > 10^{6.6} \msun$, and the right panel is its lower-resolution counterpart with the same stellar mass cut. Lowering the resolution tends to increase r$_{1/2,*}$ for a given $M_*$. However, the general trend of \texttt{ADM} satellite galaxies having smaller half-light radii compared to $\cdm$ subhalos is still present.} 
    \label{fig:res_radii}
\end{figure*}

Figure~\ref{fig:res_radii} shows the relationship between stellar mass and stellar-half-mass radius.  The left panel shows the same results as Fig.~\ref{fig:stellar_radii}, but with an increased stellar mass cut ($M_* > 10^{6.6} \msun$), while the right panel shows the corresponding result for the lower-resolution simulations. With the larger stellar mass cut, the median and 68\% containment values for $r_{1/2,*}$ in the higher-resolution $\cdm$, $\admone$, and $\admtwo$ are now $2.00_{-0.04}^{+0.24}$~kpc, $0.75_{-0.49}^{+8.71}$~kpc, and $0.42_{-0.16}^{+0.61}$~kpc, respectively. The corresponding values for the lower-resolution simulations are $5.24_{-1.06}^{+1.12}$~kpc, $1.14_{-0.35}^{+1.09}$~kpc, and $1.36_{-1.03}^{+0.42}$~kpc. There is a clear shift in the median values, beyond the observed scatter. The shift is in the expected direction, with half-light radii increasing for the lower-resolution runs. Importantly, the trend that the \texttt{ADM} simulations form more compact satellite galaxies compared to $\cdm$ is still present at lower resolution, suggesting that the qualitative trend is robust.

\begin{figure*}[t]
    \centering
    \includegraphics[width=0.49\linewidth]{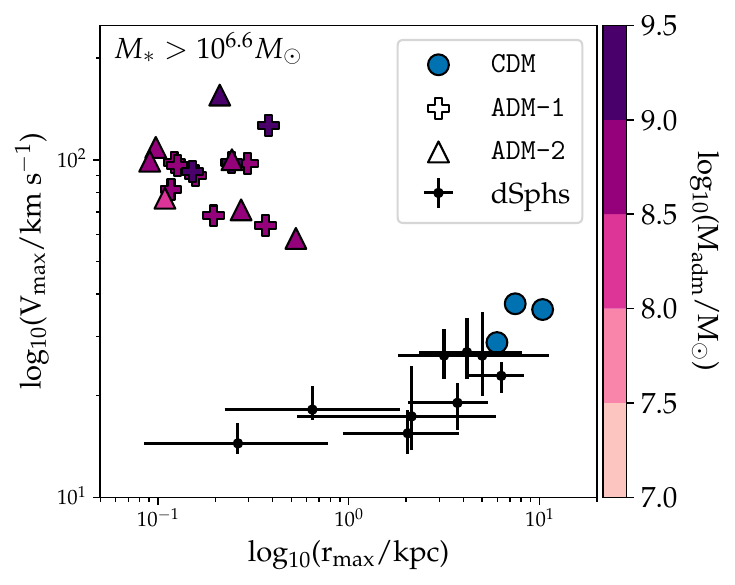}
    \includegraphics[width=0.49\linewidth]{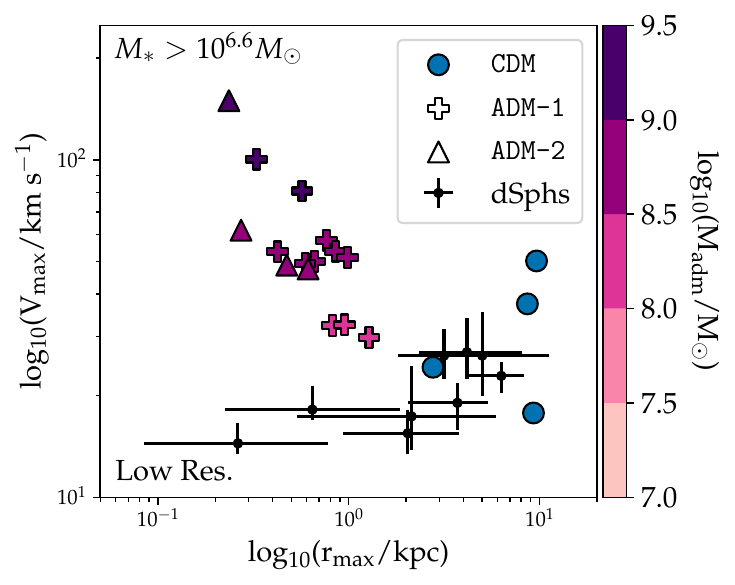}
    \caption{The left panel uses the same simulations as Fig.~\ref{fig:rmax}, but with a larger stellar mass cut, $M_* > 10^{6.6} \msun$, and the right panel is its lower-resolution counterpart with the same stellar mass cut. The $\cdm$ distribution is roughly consistent between both panels while and still exhibit the same trend discussed in the body of the paper.} 
    \label{fig:rmax_lowres}
\end{figure*}

Figure~\ref{fig:rmax_lowres} demonstrates how lowering the resolution impacts the r$_{\rm{max}} - $V$_{\rm{max}}$ distributions. With the larger stellar mass cut, the $\cdm$, $\admone$, and $\admtwo$ simulations have the following median and 68\% containment for ($r_{\rm{max}}$,$V_{\rm{max}}$): ($7.46_{-1.00}^{+1.99}$~kpc, $36.01_{-4.94}^{+0.98}$~km/s), ($0.18_{-0.05}^{+0.16}$~kpc, $94.41_{-20.25}^{+3.82}$~km/s), and ($0.21_{-0.11}^{+0.07}$~kpc, $99.01_{-28.46}^{+11.66}$ km/s). For the lower-resolution simulations, $\cdm$, $\admone$, and $\admtwo$ have median and 68\% containment for (r$_{\rm{max}}$,V$_{\rm{max}}$): ($8.97_{-3.39}^{+0.51}$~kpc$,30.86_{-9.95}^{+13.19}$~km/s), ($0.76_{-0.25}^{+0.20}$~kpc$,51.29_{-18.89}^{+15.55}$~km/s), and ($0.37_{-0.12}^{+0.17}$~kpc$,55.31_{-7.21}^{+52.17}$~km/s). While the $\cdm$ values agree within the observed scatter, the values for both \texttt{ADM} simulations shift to smaller r$_{\rm max}$ and larger V$_{\rm max}$ values as the resolution is increased. The qualitative trend that there is a clear separation between the $\cdm$ and \texttt{ADM} satellite galaxies in the r$_{\rm max}-$V$_{\rm max}$ plane is present at both resolutions. Additionally, since known resolution-dependent effects `puff up' satellite galaxies if satellite galaxies are not converged, increasing the resolution should only further increase central densities, resulting in even smaller r$_{\rm max}$ values and also larger $V_{\rm max}$ values. Thus higher-resolution simulations should not ameliorate this discrepancy between $\cdm$ and \texttt{ADM} satellite galaxies, making this qualitative trend of \texttt{ADM} having increased relative circular velocity profiles robust.

\clearpage
\section{Friends-of-Friends Linking Percentage}
\label{sec:rockstar}

\setcounter{equation}{0}
\setcounter{figure}{0} 
\setcounter{table}{0}
\renewcommand{\theequation}{B\arabic{equation}}
\renewcommand{\thefigure}{B\arabic{figure}}
\renewcommand{\thetable}{B\arabic{table}}

\texttt{ROCKSTAR} \citep{Behroozi_2012} is a 6D halo finder that works through hierarchical refinement of friends-of-friends~(FOF) groups. The FOF algorithm works by linking particles directly to other particles (friends) within a specified linking length, $l$, and then indirectly through any particles linked to its friends (friends-of-friends), forming a group. This algorithm is implemented in \texttt{ROCKSTAR} via the following steps. Initially, 3D FOF groups are established in position space from simulation particles, using a fixed linking length for easy parallelization. These groups are then individually normalized in 6D phase space by dividing particle positions and velocities by the group's position and velocity dispersions. For each of these normalized groups, a 6D linking length is adaptively chosen such that a certain percentage of particles in the group are part of a subgroup. This percentage is what is referred to as the ``FOF linking percentage.'' This process is repeated, identifying deeper substructure at each iteration, until subgroups are identified with some minimum number ($\sim$10) of particles. `Seed halos' are placed at these groups and simulation particles are assigned hierarchically to the closest seed halo in phase space.  Lastly, any unbound particles are removed. These are the subhalos identified by \texttt{ROCKSTAR} and that we study in this work.

The FOF linking percentage determines the number of particles that must be within another particle's linking radius at each iterative step. Because several simulations in our suite contain both CDM and ADM particles with different masses, we considered using a customised \texttt{ROCKSTAR} that adaptively increases the linking length until a percentage of total mass of the group is linked, rather than the number of particles. This implementation was tested and there was no difference to the identified subhalo populations using the standard version of \texttt{ROCKSTAR}.

The default FOF linking percentage used in \texttt{ROCKSTAR} for CDM simulations is 70\%. Since ADM clumps mostly occupy central regions of subhalos in large numbers, one may worry that by only requiring 70\% of particles to be linked at each subgroup, the identified subhalos will be biased towards these dense, high-occupation regions. In the main body, we conservatively increased the FOF linking percentage to 80\% to increase the number of potential subgroups we identify. In this appendix, we test the differences that arise due to changes to the FOF linking length.

\begin{figure*}[h!]
    \centering
    \includegraphics[width=0.49\linewidth]{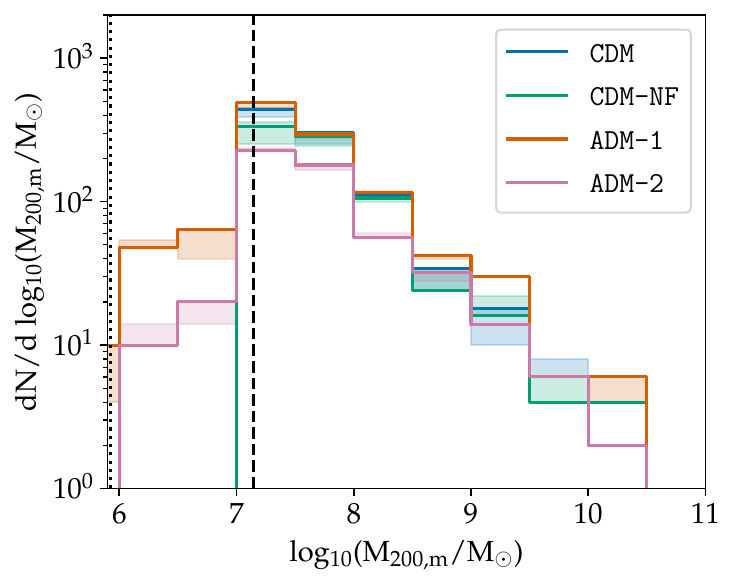}
    \includegraphics[width=0.49\linewidth]{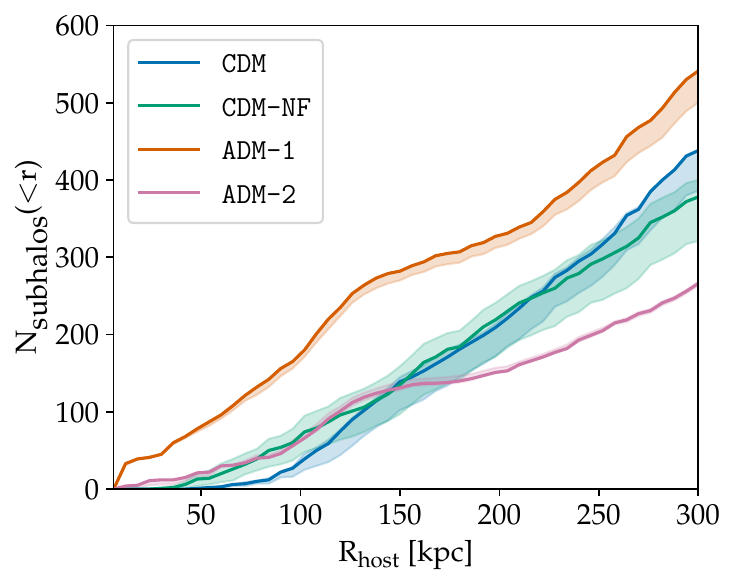}
    \caption{Mass~(left) and radial~(right) distributions for subhalos within 300~kpc of the host halo at $z=0$, similar to Fig.~\ref{fig:SHMF} and~\ref{fig:radial_dist}. In both panels, the solid line indicates the result for 80\% linking percentage, corresponding to the results shown in the body of the paper.  The  colored bands indicate the range of values when the linking percentage is decreased to 70\%~\citep{Behroozi_2012} or increased to 90\% (maximally computationally efficient value).} 
    \label{fig:rockstar_frac}
\end{figure*}

Figure~\ref{fig:rockstar_frac} shows how the subhalo mass and radial distributions for the $\cdm$, $\cdmnf$, $\admone$, and $\admtwo$ simulations vary when changing the FOF linking percentage. The solid line corresponds to a linking percentage of 80\% (used in the main body), while the bands bracket the range of values obtained when it is decreased to 70\%~\citep{Behroozi_2012} or increased to 90\% (maximally computationally efficient value). There are small quantitative changes based on the linking fraction choice, with $\cdmnf$ having the largest variation.  The overall trends---such as the plateau $\sim 125$~kpc in the $\admone$ and $\admtwo$ radial distributions---are robust, however. Additionally, increasing the linking percentage to 90\% does not result in a large increase of subhalos, confirming that most substructure is already identified at 80\% linking percentage. 

\begin{figure*}[h!]
    \centering
    \includegraphics[width=0.49\linewidth]{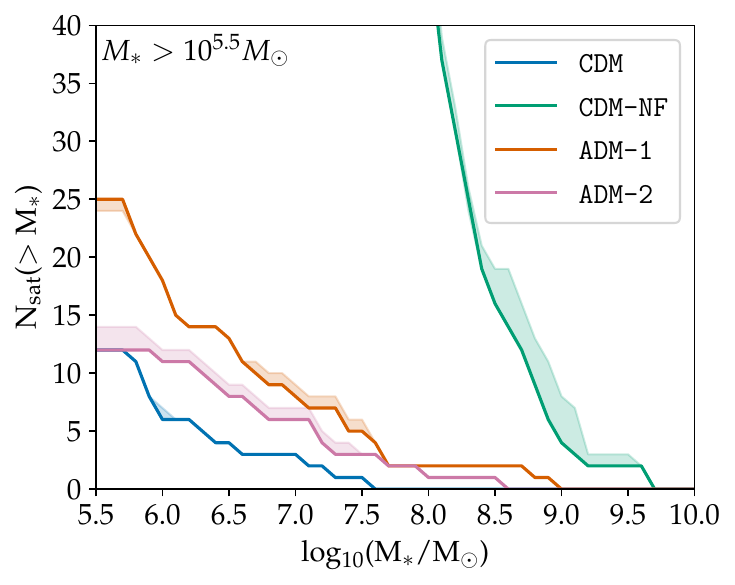}
    \includegraphics[width=0.49\linewidth]{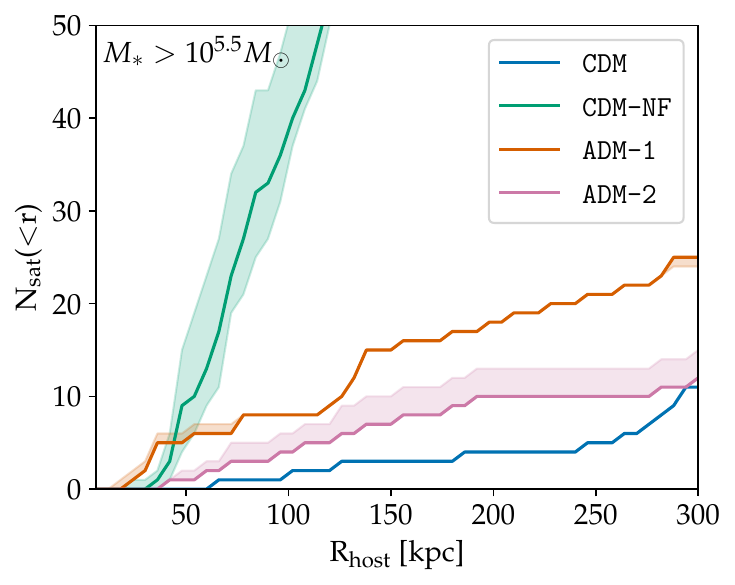}
    \caption{Mass (left) and radial (right) distributions of visible satellite galaxies within 300~kpc of the host halo at $z=0$, similar to Fig.~\ref{fig:stellar_distributions}.  In both panels, the solid line indicates the result for 80\% linking percentage, corresponding to the results shown in the body of the paper.  The  colored bands indicate the range of values when the linking percentage is decreased to 70\%~\citep{Behroozi_2012} or increased to 90\% (maximum computationally efficient value).} 
    \label{fig:rockstar_frac_sat}
\end{figure*}

Figure~\ref{fig:rockstar_frac_sat} also compares the mass and radial distributions for visible satellite galaxies using different linking percentages. Again, there is little qualitative change in the distributions, and the $\cdm$ simulation shows almost no variation across the range considered.

\clearpage
\section{Subhalo Threshold Criteria}
\label{sec:error}

\setcounter{equation}{0}
\setcounter{figure}{0} 
\setcounter{table}{0}
\renewcommand{\theequation}{C\arabic{equation}}
\renewcommand{\thefigure}{C\arabic{figure}}
\renewcommand{\thetable}{C\arabic{table}}

\begin{figure}[h!]
    \centering
    \includegraphics[width=0.5\linewidth]{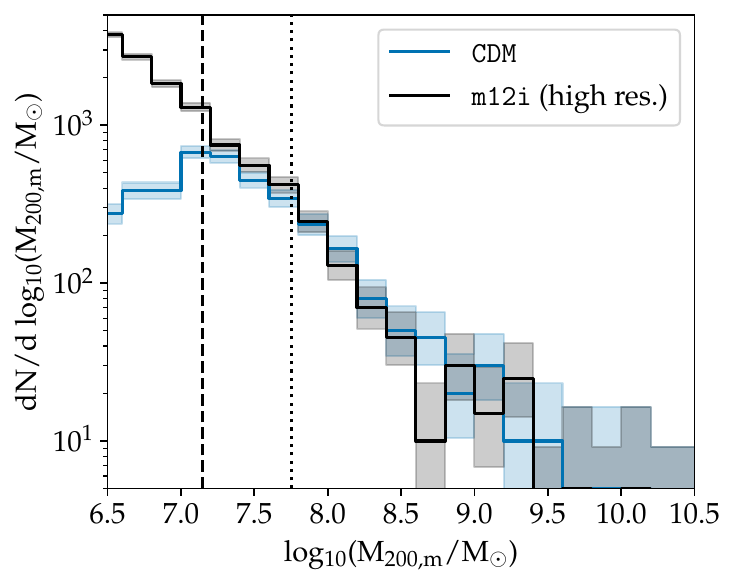}
    \caption{The mass function at $z=0$ for subhalos within 300~kpc of the host galaxy in the $\cdm$ and high-resolution \texttt{m12i} \citep{Wetzel_2023} simulations, with no cuts applied. The shaded bands indicate the 1$\sigma$ Poisson error for the count in each bin. The vertical dashed black line indicates $50 \times m_{\rm cdm} = 1.4\times10^{7} \msun$, and the dotted black line indicates $200 \times m_{\rm cdm} = 5.6\times10^{7} \msun$.}
    \label{fig:SHMF_no_cut}
\end{figure}

In the body of the paper, a value $N_{\rm{cut}} = 50$ is used to set the mass-weighted threshold. This value is chosen as it corresponds approximately to the number of particles in the $\cdm$ simulation where the subhalo mass function turns over at low masses, diverging from the the high-resolution \texttt{m12i} simulation, shown in Fig.~\ref{fig:SHMF_no_cut}. Here, we test the effect of increasing this threshold to $N_{\rm{cut}} = 200$. Note that all the visible satellite galaxies across all simulations studied satisfy this cut, so the results in Sec.~\ref{sec:sat galaxies} remain unchanged by default.

\begin{figure}
    \centering
    \includegraphics[width=0.5\linewidth]{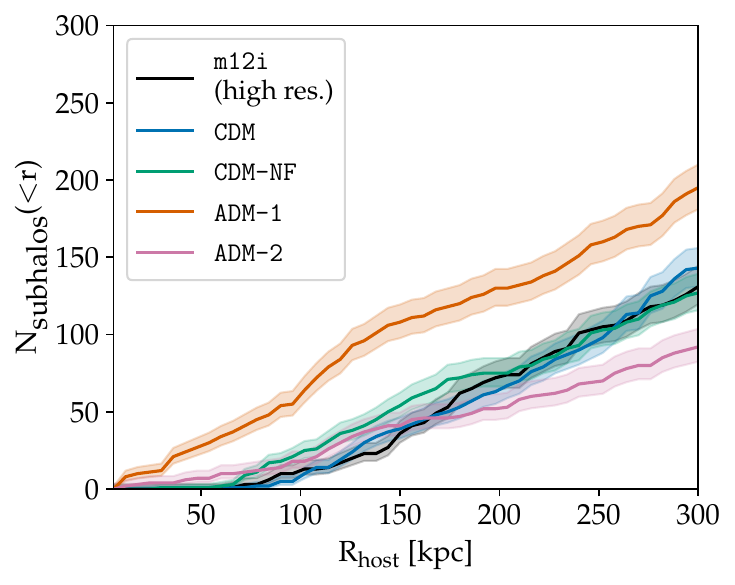}
    \caption{Cumulative number of subhalos within a given distance to the host center, at $z=0$, with a larger mass-weighted threshold ($N_{\rm{cut}}=200$).  Results are shown for the $\cdm$~(blue), $\cdmnf$~(green), $\admone$~(orange), and $\admtwo$~(pink) simulations. Shaded bands indicate the 1$\sigma$ Poisson error in the count. The solid black is the cumulative distribution obtained from the  high-resolution~(high res.) public FIRE-2 data for the \texttt{m12i} simulation \citep{Wetzel_2023}, with a mass cut $M_{200,\rm{m}} > 200 \times m_{\rm cdm} = 5.6\times10^{7} \msun$. Compared to Fig.~\ref{fig:radial_dist}, there is better agreement between $\cdm$ and \texttt{m12i} simulations, suggesting the discrepancy in the body of the paper is mainly due to low-mass subhalos. Both \texttt{ADM} simulations still contain a larger fraction of subhalos within the inner $\sim 125$~kpc of the galaxy compared to the CDM-only simulations. The total number of subhalos is still enhanced in $\admone$, but only slightly depleted in $\admtwo$.} 
    \label{fig:radial_200}
\end{figure}

Figure~\ref{fig:radial_200} shows the radial distribution of subhalos subject to $N_{\rm{cut}} = 200$; all previously identified trends remain, except with better agreement between the $\cdm$ and \texttt{m12i} simulations. The $\admone$ simulation still exhibits a relative enhancement compared to the others, while the total number of subhalos in $\admtwo$ is suppressed.  Moreover, the plateau feature in both \texttt{ADM} simulations is present from 125--200~kpc, as observed previously.

\begin{figure*}
    \centering
    \includegraphics[width=\linewidth]{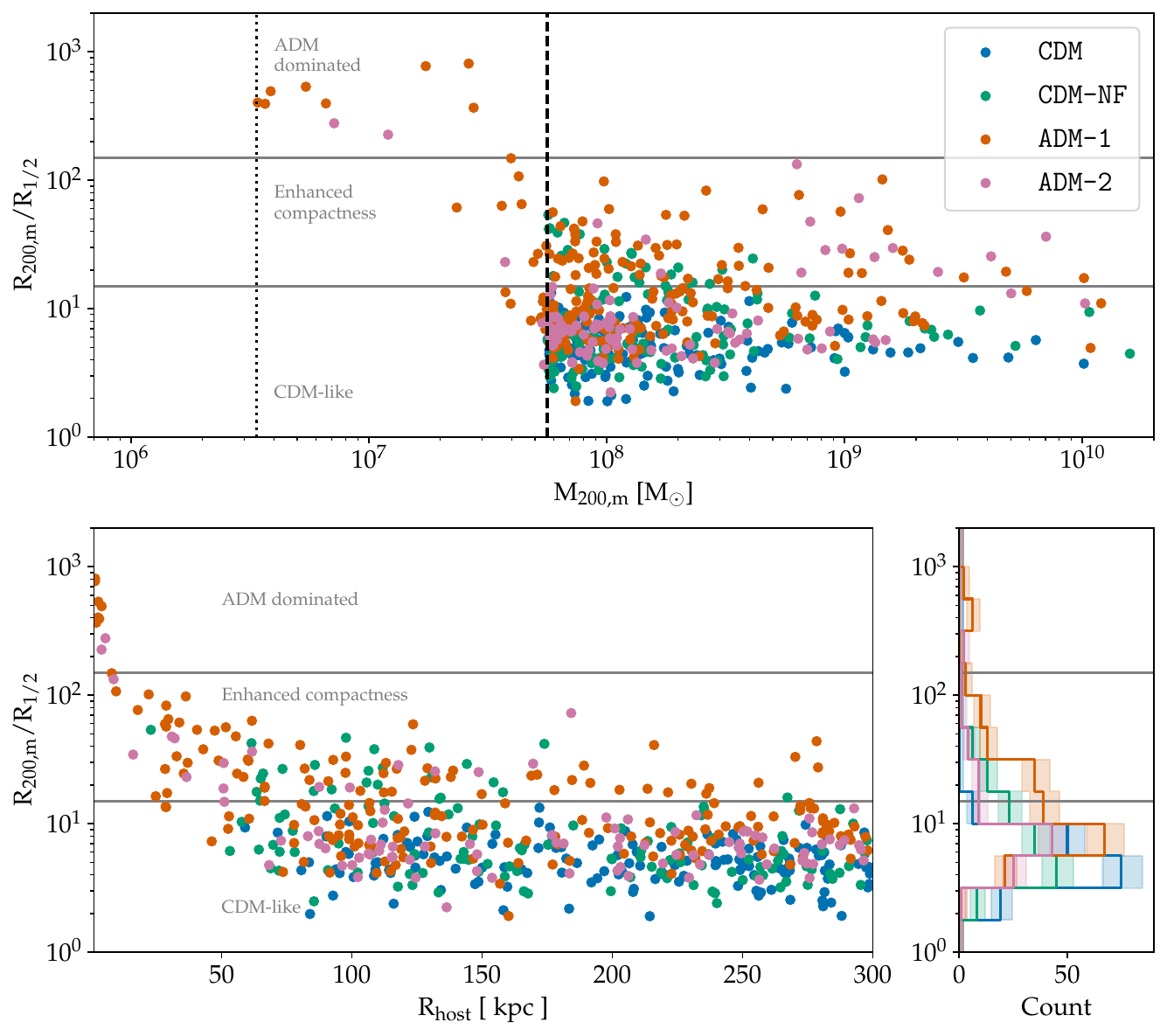}
    \caption{Distributions of subhalo compactness, as defined by $R_{200, {\rm m}}/R_{1/2}$, for $\cdm$~(blue), $\cdmnf$~(green), $\admone$~(orange), and $\admtwo$~(pink) simulations. Subhalos satisfy an increased mass-weighted threshold with $N_{\rm{cut}}=200$. \textbf{Top:} Plot of $R_{200, {\rm m}}/R_{1/2}$ versus halo mass, $M_{200, {\rm m}}$. The vertical dashed black line represents the $\cdm$ mass threshold, $200 \times m_{\rm cdm} = 5.6\times10^{7} \msun$, while the vertical dotted black line represents the \texttt{ADM} mass threshold, $200 \times m_{\rm adm} = 3.4\times10^{6} \msun$. All the previously identified and discussed groupings of subhalos are present, albeit less populated. \textbf{Bottom left:} Plot of $R_{200,{\rm m}}/R_{1/2}$ versus distance to host galaxy, $R_{\rm host}$. All trends discussed in the body of the paper are also visible even with the increased prticle number cut. \textbf{Bottom right:} Histogram of $R_{200, {\rm m}}/R_{1/2}$ values, all simulations peak in the `CDM-like' region, but the \texttt{ADM} simulations peak at slightly larger values. The Enhanced-compactness region is mainly occupied by the extended tails of the \texttt{ADM} aand $\cdmnf$ distributions. The ADM-dominated region is then mostly occupied by a small peak of $\admone$ subhalos.}
    \label{fig:concen_200}
\end{figure*}

Figure~\ref{fig:concen_200} shows the compactness of all the DM subhalos. Due to the stronger quality cut, the overall population of subhalos is smaller, but the qualitative trends identified in the body of the paper are still present. In particular, the three groups of subhalos are still present. Thus, the main conclusions regarding ADM influence on subhalo compactness are still applicable.

\end{document}